\documentclass[pre,twocolumn,superscriptaddress,showpacs]{revtex4-1}
\usepackage[breaklinks]{hyperref}
\usepackage[hyphenbreaks]{breakurl}
\usepackage{amsfonts, amssymb, amsmath}
\usepackage{graphicx}
\usepackage{color}
\usepackage{soul}

\definecolor{bulgarianrose}{rgb}{0.28, 0.02, 0.03}
 
\makeatletter
\newcommand*{\rom}[1]{\expandafter\@slowromancap\romannumeral #1@}
\makeatother

\makeatletter
\newcommand*\bigcdot{\mathpalette\bigcdot@{.5}}
\newcommand*\bigcdot@[2]{\mathbin{\vcenter{\hbox{\scalebox{#2}{$\m@th#1\bullet$}}}}}
\makeatother

\newcommand*{\E}{\mathrm{e}}
\newcommand*{\D}{\mathrm{d}}

\makeatletter
\def\Dated@name{Published: }
\makeatother
\begin{document}
\title{Transition to collective oscillations in finite Kuramoto ensembles}%Effects Of Frequency Sampling On Finite Kuramoto Ensembles}
\date{20 March 2018}
\author{Franziska Peter}
\affiliation{Institute of Physics and Astronomy, University of Potsdam, Karl-Liebknecht-Stra\ss{}e 24-25, 14476 Potsdam, Germany}
\author{Arkady Pikovsky}
\affiliation{Institute of Physics and Astronomy, University of Potsdam, Karl-Liebknecht-Stra\ss{}e 24-25, 14476 Potsdam, Germany}
\affiliation{Research Institute for Supercomputing, Nizhny Novgorod State University,
Gagarin Av.\ 23, 606950, Nizhny Novgorod, Russia}
\begin{abstract}
  We present an alternative approach to finite-size effects around the synchronization transition in the standard Kuramoto model. Our main focus lies on the conditions under which a
  collective oscillatory mode is well defined. For this purpose, the minimal value of the amplitude of the complex Kuramoto order parameter appears as a proper indicator.
  The dependence of this minimum on coupling strength varies due to sampling variations and 
  correlates with the sample kurtosis of the natural frequency distribution. 
  The skewness of the frequency sample determines the frequency of the resulting collective mode. The effects of kurtosis and skewness hold in the thermodynamic limit of infinite ensembles. We prove this by integrating a self-consistency equation for the complex Kuramoto order parameter 
  for two families of distributions with controlled kurtosis and skewness, respectively.
  %For the infinite Gauss- and Laplace distribution, the self-consistency approach gives an parametric solution for the   order parameter expressed by Bessel functions.
\end{abstract}

\maketitle

\section{Introduction}

Synchronization in ensembles of self-sustained oscillators is a universal phenomenon, relevant not only for many physical and technical applications (e.g.,
laser arrays~\cite{Nixon_etal-13}, electrochemical oscillators~\cite{Kiss2002}, 
and power grids~\cite{Motter-13}) but also for the self-perpetuation of living beings. 
In such biological systems, the manifestations of synchrony are often spectacular--like the united 
rhythmic flashing of mating fireflies that attracts tourists over vast distances or synchronized brain 
waves that apparently speed up learning~\cite{Miller2014}.  
From single-cell organisms to animals through to humans, many species benefit from synchronizing 
their motions, metabolism, cell division, gene expression, circadian cycles
and many other inner rhythms.

Arthur Taylor Winfree's model of large ensembles of self-sustained 
oscillatory systems~\cite{Winfree1980} and its modification to an analytically solvable 
mean-field system by
 Yoshiki Kuramoto~\cite{Kuramoto1975, Kuramoto1984} established the understanding of synchronization as a
 nonequilibrium phase transition. 
Since then, various extensions to this Kuramoto model contributed to a broader, 
more realistic, picture of 
synchronization dynamics in large ensembles: the generalization to higher 
modes of the coupling 
function~\cite{Daido-96a,Komarov2014}, the introduction of a phase 
shift~\cite{Sakaguchi_Kuramoto86} 
to the coupling 
function, the addition (or multiplication) of noise~\cite{Acebron2005}, and 
the investigation of different coupling network 
structures~\cite{Arenas2008}. Two significant analytical achievements 
allow for a deeper 
understanding of the phase space geometry: the Watanabe-Strogatz reduction 
for finite ensembles of identical 
oscillators~\cite{Watanabe1994} and the Ott-Antonsen \textit{ansatz} for infinite 
ensembles of distributed 
oscillators~\cite{Ott2008}.

Most of the present theoretical approaches to synchronization concentrate on infinite populations, 
while the theory on finite-size ensembles evolves only gradually. However, only a few experiments
 study really large numbers (thousands) of oscillatory units~\cite{Prindle_etal-12,Nixon_etal-13}, 
 while experimental setups of
 populations up to a hundred oscillators dominate the area~\cite{Kiss2002,Hauser2012, PhysRevE.85.015204}. 
Many numerical and theoretical studies focus on the properties of either rather small systems--e.g.,  Ref.~\cite{Popovych-Maistrenko-Tass-05} explores the
chaotic dynamics in a Kuramoto model with only $4$ to $20$ oscillators--or much larger 
ensembles of $10^4$ or more oscillators, where most efforts are dedicated to the scaling 
properties of fluctuations of the order parameter~\cite{Daido1987, Daido1990, Son2010, Hong2007, Hong2015}. 

In this paper we explore ensembles of moderate size (typically 50 to 200 oscillators)--corresponding to realistic 
experimental conditions. In this respect, this study contributes to closing 
the gap between the two extremes of relatively small and very large ensembles
by investigating effects that naturally emerge in small ensembles and extend
to the infinite limit depending on the natural frequency distribution.

Our approach to the finite-size problem differs from the scaling-of-fluctuations approach 
adopted in~\cite{Daido1987, Daido1990, Son2010, Hong2007, Hong2015}. We dedicate 
the primary focus to the question: Under which 
conditions is a collective mode well defined in a finite ensemble of coupled phase oscillators? In the thermodynamic limit, 
the exact dynamical equations for the complex order parameter can be derived in some cases.
The most prominent example is the Ott-Antonsen theory for the Kuramoto-Sakaguchi system~\cite{Ott2008}.
In other cases, at least an asymptotic solution for infinitely large ensembles can 
be interpreted as oscillations of a collective mode. 
In general, the main feature that distinguishes
self-sustained oscillations from the noise-driven ones is the existence of a macroscopic phase: It is well defined
for self-sustained oscillations, but ill defined for noisy states where the amplitude can vanish.

In finite ensembles of Kuramoto type--in contrast to the thermodynamic limit--the complex order parameter fluctuates strongly.
 It is suggestive to consider a collective mode as a well-defined macroscopic oscillation, 
if the corresponding macroscopic phase is well-defined for all times. This means that the amplitude
should not vanish (cf. Ref.~\cite{PhysRevE.85.015204} where this idea applies to experimental
studies of a finite set of oscillators). Below, we study in detail the statistical properties of the minimum of
the amplitude of the complex order parameter that serves as an indicator for the emergence of a
global oscillatory mode. These properties strongly depend on the particular sample of  frequencies.
The observed effects can be traced in the thermodynamic limit by exploring distributions with controlled
kurtosis and skewness.

The paper is organized as follows: In Sec.~\ref{sec:Rmin_bifu}, we introduce the model and 
discuss the phenomenology of the complex order parameter dynamics for both infinite and finite ensembles. 
Section~\ref{sec:minR} introduces the minimum of the amplitude 
of the order parameter as an indicator for the presence of a collective mode. 
Section~\ref{sec:order_kurt} discusses the effect of sample kurtosis and sample
skewness of the natural frequency distribution on the synchronization transition characterized by
the introduced indicator and on the global phase dynamics, respectively. 
These properties persist in the thermodynamic limit, as we prove in Sec.~\ref{sec:samp_dist} for distribution families
with parameters for kurtosis and skewness, respectively. 
Section~\ref{sec:concl} contains conclusion and outlook.

 \section{Synchronization transition: thermodynamic limit vs. finite system}\label{sec:Rmin_bifu}%{The Kuramoto Order Parameter: Transition to a Collective Phase} 
In this section, we introduce the Kuramoto model of coupled phase oscillators and compare the dynamics 
of the complex order parameter in infinite and finite ensembles. We discuss the role of coupling strength and 
natural frequencies in both cases.

The standard Kuramoto model describes $N$ nearly identical phase oscillators with weak sinusoidal coupling.
Their natural frequencies $\omega_i$ spread according to some distribution $g(\omega)$.
Phases $\theta_i$ are globally coupled with strength $\epsilon$, 
\begin{equation}\label{eq:kura}
  \dot\theta_i = \omega_i + \frac{\epsilon}{N}\,\sum^N_{j=1}\sin(\theta_j-\theta_i) = \omega_i + \epsilon R\,\sin(\varphi-\theta_i), 
 \end{equation}
via a complex mean field $Z$, defined as
 \begin{equation}\label{eq:order}
 Z = R\,\mathrm{e}^{\imath\varphi} = \frac{1}{N}\,\sum^N_{j=1}\,\mathrm{e}^{\imath\theta_j}.
 \end{equation}
The absolute value of $Z$, called Kuramoto order parameter $R$, quantifies the degree of 
phase coherence in the population and thereby serves as an indicator for synchrony.

Shifting all frequencies by a constant $\omega_i \rightarrow \omega_i+\Delta\omega$ is 
equivalent to transforming to a 
rotating reference frame with frequency $\Delta\omega$ to the entire system. This rotational invariance 
proves beneficial in the thermodynamic limit, where the complex mean field rotates 
uniformly. This rotation becomes stationary in an appropriate reference frame, which us allows us to describe a synchronous state
as a steady one.
Similarly, scaling all frequencies by a constant 
factor $\omega_i \rightarrow \sigma\omega_i$ just scales time and the coupling strength 
by the same factor. Thus, without loss of generality, we set the standard deviation 
of $g(\omega)$ to 1 and shift the mean frequency to zero in all examples and all 
numerical experiments (in the latter case: after sampling).

\subsection{Solution of the Kuramoto model in the thermodynamic limit}
\label{sec:kurt}

 % Applying the limit of infinitely many oscillators and assuming a Lorentzian natural frequency distribution, Kuramoto solved the mean field dynamics in a self-consistent way. 
 
 %[1p] critical coupling, second order transition, from incoherence to synch, different distributions.
 In the thermodynamic limit of infinite ensembles, $N\rightarrow\infty$, the 
 dynamics of the complex mean field $Z$ as a function of coupling strength $\epsilon$ 
 demonstrates a transition to synchrony -- comparable to a nonequilibrium phase 
 transition \cite{Kuramoto1975, Kuramoto1984}.
 In this section we recall a basic qualitative picture of this transition for symmetric unimodal
 distributions (thus assuming their maximum at the mean frequency $\bar\omega$) 
 and describe a quantitative method for finding the order parameter as a function of $\epsilon$.
 
States $\{\theta_i\}$ with vanishing order parameter are always solutions, irrespective of the coupling 
strength. With $R=0$, the oscillators perfectly decouple and rotate 
with their respective natural frequencies. Therefore, the individual phases are fully 
incoherent, i.e., uniformly distributed in $[0, 2\pi)$, so that $R$ vanishes exactly, reflecting the self-consistent nature of the problem.
Another, nontrivial, solution with $R>0$ exists above the critical coupling $\epsilon_c^\infty = 2\cdot[\pi g(\bar\omega)]^{-1}$. 
The order parameter $R$ as a function of coupling strength $\epsilon$ is typically continuous but not differentiable in $\epsilon_c^\infty$. 
Only distributions with a symmetric plateau around $\bar\omega$ produce a jump at the critical coupling strength~\cite{Basnarkov2007}, with
the uniform distribution as a special case~\cite{Pazo2005}. (Multimodal 
distributions typically exhibit hysteresis~\cite{PhysRevE.80.046215}.)

Coupling strengths above $\epsilon_c^\infty$ may lock only a fraction of 
 the oscillators to a common frequency--except, e.g., in the case of a uniform frequency 
 distribution, where the oscillators jump from zero to full frequency locking at $\epsilon_c^\infty$. The fraction of frequency locked oscillators increases with coupling strength, reflected in a growing order parameter. For distributions 
 with compact support, the maximal frequency difference determines a coupling strength above which
 all oscillators rotate with the same observed frequency. As long as $\epsilon$ is 
 finite, they maintain finite phase 
 differences, and the order parameter asymptotically approaches $R=1$. For 
 distributions with unbounded support,
 the fraction of asynchronous oscillators is always finite.
 
 Quantitatively, the Kuramoto problem in the thermodynamic limit can be solved as 
 follows~\cite{Omelchenko-Wolfrum-12,Omelchenko-Wolfrum-13,ZhangPikovskyLiu2017}.
 One seeks for a solution that is stationary (in the sense of a stationary distribution function of phases, see Ref.~\cite{Ott2008})
 in a frame rotating with some frequency $\Omega$, i.e., $\varphi=\Omega t+\varphi_0$, where $\varphi_0$ is a constant.
 The relative phase $\psi=\theta-\Omega t-\varphi_0$ obeys 
 \begin{equation}\label{eq:kura2}
  \dot\psi = \omega -\Omega - \epsilon R\,\sin(\psi)\;,\qquad R=\langle e^{i\psi}\rangle. 
 \end{equation}
It is convenient to consider $\Omega$ and $a=\epsilon R$ as parameters in Eq.~\eqref{eq:kura2}. The distribution
of phases $\psi$ at given $\omega$ can then be expressed as 
$\delta(\psi-\arcsin[(\omega-\Omega)/a)]$ for synchronous
 and $\sqrt{(\omega-\Omega)^2-a^2} |\omega-\Omega-a\sin\psi|^{-1}$ for asynchronous oscillators, 
 respectively. The definition of the mean field $R$ can be expressed as a complex function of $\Omega$ and $a$, consisting of one real and two complex integrals (the other real integral, stemming from the asynchronous oscillators combined with the cosine, vanishes due to $2\pi$ periodicity):
 \begin{equation}\label{eq:kura3}
 \begin{gathered}
 R=F(\Omega,a)  = a\, \int\limits_{-\pi/2}^{\pi/2}\,\E^{i \theta}  g(\Omega +a\sin\theta)\cos\theta\,\D\theta \\
                 +  \frac{i}{2\pi}\int\limits_{-\pi}^\pi\,\,\sin \theta \int\limits_{|\omega-\Omega|> a} g(\omega)\, \frac{\sqrt{(\omega-\Omega)^2-a^2}}{|\omega-\Omega-a\sin\theta|}\,\D\omega\,\D\theta .
\end{gathered}
\end{equation}

 In practice, both analytical and numerical integration of this equation provide a solution 
 in parametric form with parameter $a$. First, as $R$ is a real quantity, the imaginary part 
 of the right-hand side must vanish. This condition assigns a unique value $\Omega_a$ to 
 frequency $\Omega$ for given $a$. For symmetric distributions, the imaginary integral vanishes and $\Omega = 0$. The remaining real integral $F(\Omega_a,a)$ provides the 
 dependence of order parameter $R$ on coupling strength $\epsilon$, namely 
 $R=F(\Omega_a,a)$ and $\epsilon=a/F(\Omega_a,a)$.
 
 For some special frequency distributions $g(\omega)$, the integrals in Eq.~\eqref{eq:kura3} can be calculated 
 analytically. 
 The simplest case of identical oscillators [such that $g(\omega)$ degenerates to a delta distribution] has only one 
 nontrivial solution,  $R= 1$. For a Gaussian distribution $g(\omega) = \mathrm{e}^{-\omega^2/2}/(\sqrt{2\pi})$, the 
 parametric solution can be expressed via the modified Bessel functions of the first kind 
 $\mathcal{I}_\mu$~\cite{Abramowitz}:
 \begin{equation}
  R =  \sqrt{\pi A/2}\,\mathrm{e}^{-A} \,\left[\mathcal{I}_0\left(A\right)+\mathcal{I}_1\left(A\right)\right],
 \end{equation}
 where $A= a^2/4$. Similarly, the solution for the Laplace distribution 
 $g(\omega)=\mathrm{e}^{-\sqrt{2}|x|}/(\sqrt{2})$ 
 includes the modified Bessel functions of the first kind $\mathcal{I}_\mu$ and the modified Struve functions $\mathcal{L}_\mu$: 
 \begin{equation}
  R = \pi\cdot\left[\mathcal{I}_1\left(B\right)-\mathcal{L}_1\left(B\right)\right]/2,
 \end{equation}
 where $B=a\sqrt{2}$. 
 
 For uniform distributions with mean zero, height $h$, and width $2\omega_\mathrm{max}$ (with normalization $2\omega_\mathrm{max}\cdot h=1$), solutions with $\omega_\mathrm{max}/a \ge 1$ correspond to $\epsilon_c^\infty = 2/(\pi h)$ with $R$ in $(0, \pi/4]$. Solutions with $\omega_\mathrm{max}/a < 1$ obey the parametric equation
 \begin{equation}\label{eq:sce_uniform}
  R = 2a\,\omega_\mathrm{max}\,\arcsin\left(\frac{\omega_\mathrm{max}}{a}\right) + \frac{1}{2}\sqrt{1-\left(\frac{\omega_\mathrm{max}}{a}\right)^2}.
 \end{equation}
 Due to Eq.~\eqref{eq:kura}, oscillators with $|\omega_i|<\epsilon R$ have a stable fixed point (i.e., lock to the frequency of the global phase), which is true for any coupling stronger than $\epsilon_c^\infty=2/(\pi h)$, where $R>\pi/4$, such that $\epsilon R>1/(2 h) =\omega_\mathrm{max}\ge|\omega_i|$.
This means that at the critical coupling strength all oscillators jump to full synchrony in the sense of full frequency locking.

 A number of other distributions can be integrated by the same method as well, 
 for instance, the distributions listed in Ref.~\cite{Pazo2005}.
 
\begin{figure*}[ht!]
\centering
 \includegraphics[width = \textwidth]{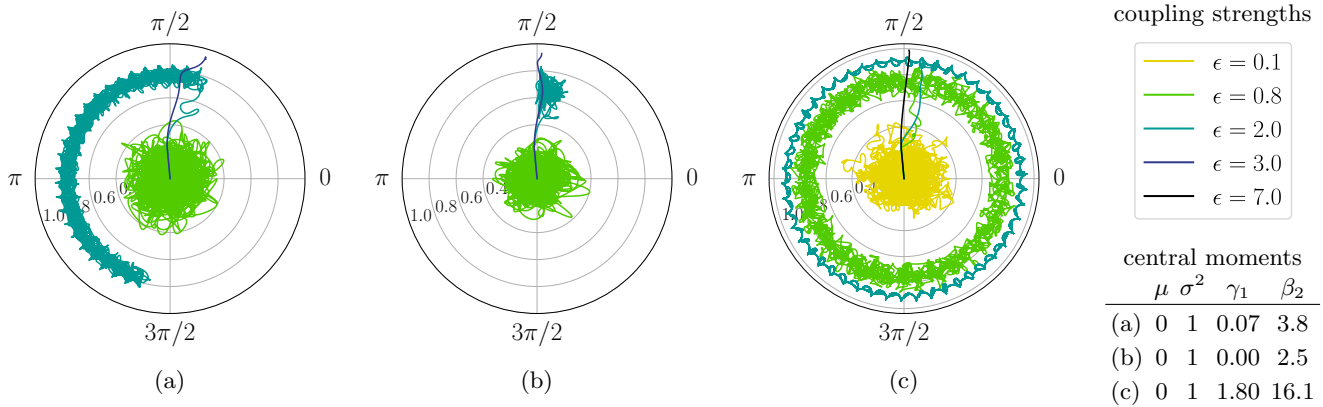}
 \caption{(Color online) Polar representation of the time evolution ($t =10^3$)  of the complex order 
 parameter $Z$  at different values of coupling strengths $\epsilon$ for $50$ oscillators with frequencies: 
 (a) randomly sampled from a normal distribution, (b) regularly sampled via quantiles from a normal distribution, 
and (c) randomly sampled from the Lorentz distribution. Note that in panels (a) and (b)
we plot only $\epsilon= 0.8,$ $2.0$, and $3.0$, while in (c) 
we show $\epsilon= 0.1$, $0.8$, $2.0$, and $7.0$  for better clarity. The table 
 shows the first sample moments of the three natural frequency samples: mean $\mu$, 
 variance $\sigma^2$, skewness $\gamma_1$, and kurtosis $\beta_2$. 
 All trajectories start from the same initial phases, randomly picked 
 from a uniform distribution in $[0,2\pi)$, such that $R(t_0)\approx 0$.  
 All numerical time evolutions in this paper use the fourth-order 
 Runge-Kutta scheme with step size 0.01.} 
  \label{fig:cmplx_order_time}
 \end{figure*}
 
\subsection{Transition and phenomenology in finite populations}

In a finite population, the solutions described above are not exact, most obviously evident in the 
fluctuations of the order parameters $Z$ and $R$. 
These fluctuations start from a finite value $\sim 1/\sqrt{N}$ for vanishing coupling, because states 
with $R=0$ are not invariant in the 
finite model. Fluctuations increase around the critical coupling and then decay for stronger coupling. 
Thus, the transition to synchrony becomes blurry when measured by means of the averaged order 
parameter. The details of the dynamics of the order parameters strongly depend on the way the finite frequency 
sample 
is generated from the underlying distribution $g(\omega)$. Here, either regular or random sampling may be 
implemented, depending on the specific question. Regular sampling by virtue of the 
quantiles of the distribution (i.e., inverse transform sampling from equidistant points)
allows for a straightforward comparison between different ensemble sizes, as in this case the frequencies in each 
sample are uniquely determined by the 
system size $N$. Random sampling (i.e., sampling with some random number 
generator) results in a finite sample-to-sample variability which decreases as $N$ grows. This variability of 
samples complicates comparisons, as a statistical analysis becomes necessary. Yet, it is the method of choice to fully represent 
the finite-size effects of the underlying distribution in an unbiased fashion.

Figure~\ref{fig:cmplx_order_time} shows a typical time evolution of the complex order parameter in an 
 ensemble of $50$ oscillators for three samples of $g(\omega)$. In the two left panels, a normal 
 distribution $\mathcal{N}(0,1)$ is sampled (a) randomly and (b) regularly. Figure~\ref{fig:cmplx_order_time}(c) 
 shows $Z(t)$ for a random sample of a Lorentz distribution which is widely used in studies on the 
 Kuramoto model because of analytic tractability in the infinite limit.
 In all three cases, the order parameter $Z$ fluctuates around zero for small coupling strengths, 
 corresponding to predominantly asynchronous motion below the synchronization transition. 
 For stronger coupling $\epsilon$, the complex order parameter $Z=Re^{i\varphi}$ is clearly 
 separated from zero.
 For moderate coupling strengths and regular sampling, only the amplitude $R$ performs sustained fluctuations, 
 while the argument $\varphi$ converges to a constant (see Secs.~\ref{sec:drift} and \ref{sec:skew_infty} for an explanation). In contrast, random samples exhibit sustained fluctuations 
 in both $R$ and $\varphi$, for the same moderate coupling strengths. In all cases, $Z$ quickly converges to a 
 complex constant for sufficiently strong coupling. The coupling strength necessary to achieve this state is 
 considerably higher ($\epsilon \sim 7$) for the Lorentz distribution sample than for the two Gaussian samples ($\epsilon \sim 3$).
 
Most of the previous works on finite-size effects in the Kuramoto 
transition~\cite{Daido1987, Daido1990, Hong2007a, Hong2015, Bronski2012} focus
on the statistics of fluctuations of the real order parameter $R$. Let us here focus on the
phase of the complex mean field $\varphi$. This is important if we want to interpret the dynamics
of the complex mean field $Z$ as that of a complex amplitude of an effective collective oscillatory mode.
As mentioned briefly in the Introduction, such interpretation is mathematically justified for the Kuramoto model in the thermodynamic
limit with Lorentz-distributed frequencies. In this case, the Ott-Antonsen ansatz~\cite{Ott2008} reduces
the dynamics to a Stuart-Landau-type equation for the order parameter $Z$. In other cases, such a reduction
is justified at least close to the transition point~\cite{Chiba-Nishikawa-11,Dietert-16}.
From this macroscopic viewpoint, the transition to synchrony in a population corresponds to a Hopf 
bifurcation from a fixed point to stable self-sustained oscillations of the mean field. The dynamics of the 
macroscopic phase $\varphi$ displays the temporal coherence of these oscillations. 
 
  In all three cases depicted in Fig.~\ref{fig:cmplx_order_time}, the order parameter 
  strays around zero for small coupling strengths but avoids an inner circle for stronger coupling. This implies a 
  problem in the definition of a macroscopic phase $\varphi$ for weak coupling: If amplitude $R$ vanishes, then the phase
  of oscillations is ill defined. This suggests to distinguish the two domains of the dynamics of the complex mean field,
  according to the minimal possible value of the amplitude $R_\mathrm{min}$ (see 
  Ref.~\cite{PhysRevE.85.015204}, where this parameter
  has been applied to the analysis of experiments with a finite set of coupled oscillators):
  \begin{itemize}
  \item If $R_\mathrm{min}=0$, then the macroscopic phase $\varphi$ is not defined globally, and thus macroscopic oscillations are ill defined. The complex order parameter diffuses around zero.
  \item If $R_\mathrm{min}>0$, then the macroscopic phase $\varphi$ is well defined. 
  It validates the term ``macroscopic oscillations'' and defines their coherence. 
  The complex order parameter is well separated from zero.
  \end{itemize}
  In the following section, we focus on the properties of $R_\mathrm{min}$.

\begin{figure}[!htb]
 \centering
   		\includegraphics[width = 0.9\columnwidth]{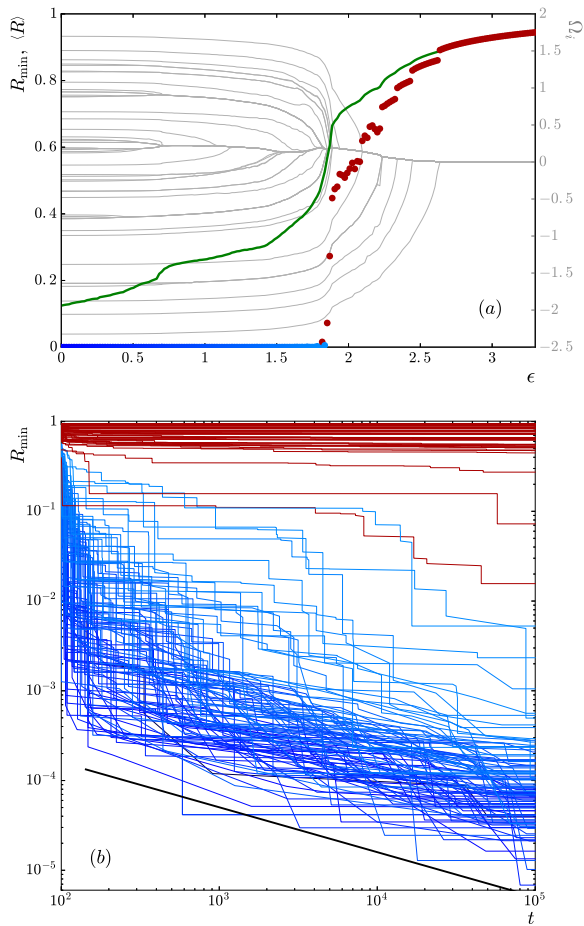}		
   \caption{(color online) 
   Statistical characterization of the minimum of $R(t)$,
for $N=50$ oscillators and a fixed random sample of a Gaussian frequency distribution 
with sample moments $\mu =0$, $\sigma^2=1$, $\gamma_1=-0.45$, $\gamma_2=-0.55$. 
Time evolution starts from uniformly distributed phases in $[0, 2\pi)$.
(a) $R_\mathrm{min}$, $\langle R\rangle$, and $\Omega_i$ vs. coupling strength $\epsilon$.
Here $R_\mathrm{min}$ is shown vs. $\epsilon$ at $t=10^5$ (after transients of length $10^2$) for an $\epsilon$ grid
  with $\Delta\epsilon\approx0.02$. The green bold solid line shows the mean 
  value $\left<R\right>_T$ averaged over a time interval of $T = 10^4$. The observed individual 
  frequencies $\Omega_i$ shown in gray (right-hand scale) reveal which oscillators 
  synchronize already at small frequencies and which join the synchronous cluster
  only at stronger coupling. The sampled natural frequencies $\omega_i$ equal the observed 
  frequencies $\Omega_i$ at zero coupling.
  (b) Dependence of $R_\mathrm{min}$ on the observation time. 
The bold black line gives an estimate $t^{-1/2}$ for the scaling behavior. In both panels, red indicates
$R_\mathrm{min}> 0.01$ at $t = 10^5$, while we color the remaining sub-critical trajectories in blue. 
   }
   \label{fig:rmin}
  \end{figure}

\section{Minimal value  of the order parameter as an indicator 
for the transition to a collective mode}
\label{sec:minR}

In the preceding section, we argued that $R_\mathrm{min}$ is an appropriate quantity
to characterize the emergence of collective oscillations in a finite population. In this section, we discuss in detail
the statistical properties of this parameter. In Fig.~\ref{fig:rmin}, we show the dependence of $R_\mathrm{min}$,
calculated over a time interval $0\leq t\leq 10^5$, on the coupling parameter $\epsilon$
for one random sample of a Gaussian distribution of frequencies. In contrast to time-averaged value 
$\langle R\rangle_t$ which smoothly depends on $\epsilon$, $R_\mathrm{min}$ undergoes a 
sharp transition at $\epsilon_c^\mathrm{min}\approx 1.82$.

From a statistical point of view, the calculation of  $R_\mathrm{min}$ is less stable than that of the averaged value, as
it is dominated by the tail of the distribution of $R$. This is illustrated in
Fig.~\ref{fig:rmin}(b), where we show $R_\mathrm{min}$ as a function of observation time.
For strong coupling, $R_\mathrm{min}$ saturates already at about $t\approx 10^3$. By contrast, 
at weak or even vanishing coupling, where the 
oscillators effectively decouple, $R_\mathrm{min}$ has no lower bound. Due to their different 
frequencies, a vicinity of any configuration of 
phases is visited and that vicinity shrinks with growing length of the time series. Thereby, 
arbitrary small values of $R_\mathrm{min}$ become increasingly probable with longer 
observation time. The finite-time observation roughly 
follows the law $R_\mathrm{min}\sim t^{-1/2}$ (black bold line), which compares to a 
random sampling of a two-dimensional distribution of $R$ with finite density at zero. 

The numeric evaluation of $R_\mathrm{min}$ is most unreliable near the critical point, where 
the decrease of $R_\mathrm{min}(t)$ with observation time $t$ 
is extremely slow (one can see several such realizations
in Fig.~\ref{fig:rmin}). This appears unavoidable, because the time 
scale typically diverges at the criticality, see, e.g., Ref.~\cite{Choi2013}.  
Nevertheless, the sharp transition in the 
dependence of $R_\mathrm{min}$ on $\epsilon$ is well pronounced and 
reliable for calculations, see Fig.~\ref{fig:rmin}(a). 

Beyond transition, $R_\mathrm{min}(\epsilon)$ follows a curve that is generally growing, but not 
everywhere monotonous. At even stronger coupling, several seemingly quite regular jumps dominate the 
picture. To understand this, a juxtaposition of $R_\mathrm{min}(\epsilon)$ and the 
individual observed frequencies $\Omega_i$ of the 
oscillators (in gray) is quite instructive: The jumps correspond to events where oscillators join the major synchronous cluster, built 
by oscillators with similar natural frequencies. At these jumps, the number of incommensurate contributions 
to $Z$ decreases and the dynamics of $Z(t)$ becomes more ordered, and eventually periodic when all 
oscillators have joined the synchronized cluster.

Remarkably, the oscillators' observed frequencies reveal frequency-locked clusters far below the critical coupling. 
The vast majority of oscillators, however, joins the central synchronized cluster at about the critical coupling strength. 

In Fig.~\ref{fig:rmin}, we present the dependence of $R_\mathrm{min}$ on coupling 
strength and observation time for just one random sample of a Gaussian distribution [see Fig.~\ref{fig:rmin}(a) at $\epsilon=0$]. 
The next section discusses the sampling variation
of the observed effects and the scaling with ensemble size $N$.

\section{Effects of Kurtosis and Skewness in Finite Ensembles}\label{sec:order_kurt}
The moments of a finite random sample are random variables, distribution
of which depends on $N$ and on the 
underlying distribution. For instance, the mean of a Gaussian sample 
of size $N$ of a Gaussian with variance $\sigma^2$ is itself Gaussian distributed 
with variance $\sigma^2/N$. In Eq.~\eqref{eq:kura}, changing mean and 
variance of $g(\omega)$ merely corresponds to shifting to a different rotating 
reference frame and a different coupling parameter range, respectively. The deviations of the next 
higher moments, skewness and kurtosis, in contrast, cannot be rescaled and 
potentially determine properties of the transition and the dynamics in finite ensembles. 
In this section, we investigate the effect of sample skewness and 
sample kurtosis of a Gaussian natural frequency distribution. 
 \begin{figure*}
   \centering
  \includegraphics[width=\textwidth]{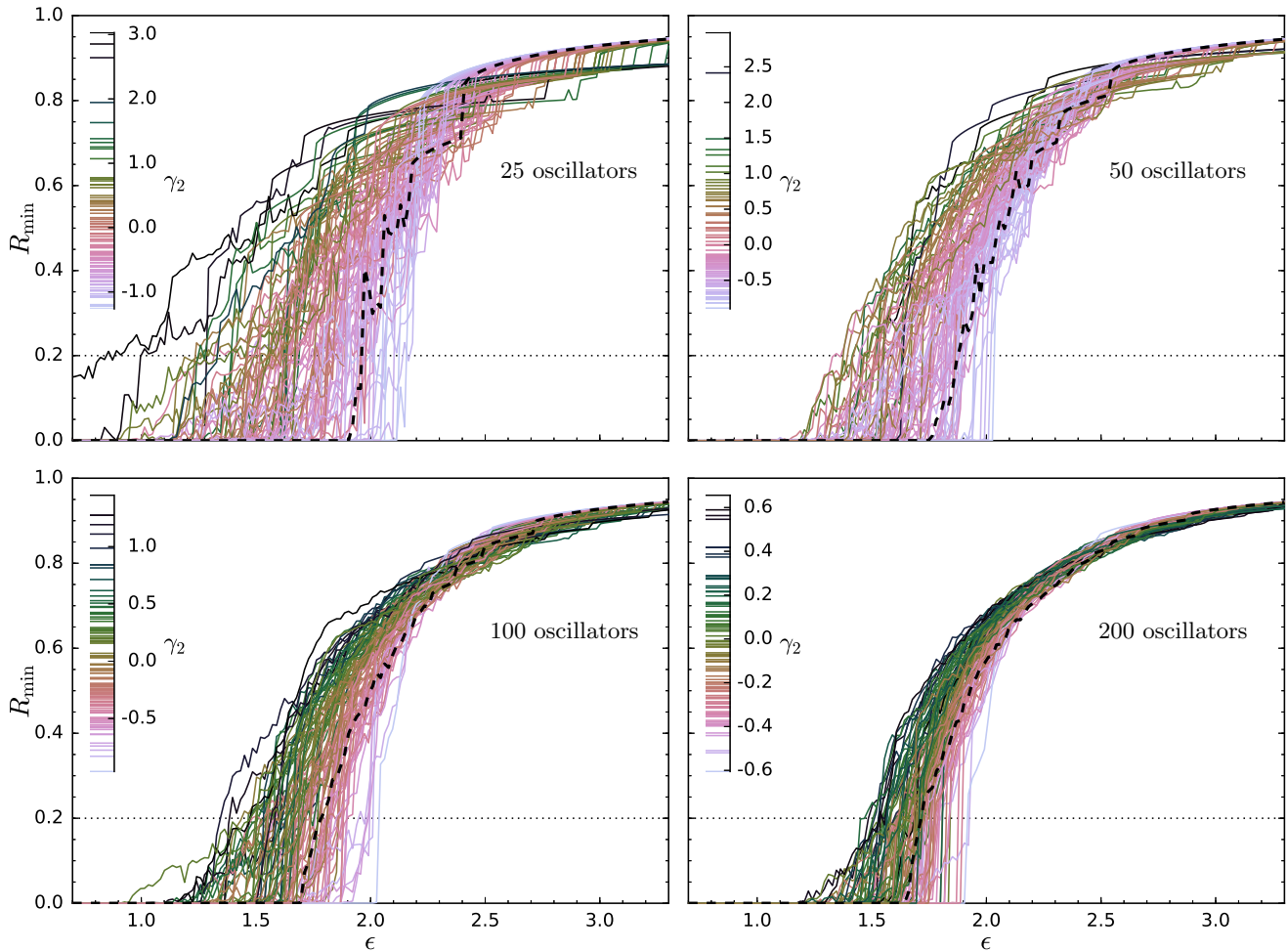}
       \caption{(color online) Dependence of $R_\mathrm{min}$ on $\epsilon$: Each panel shows results for 
   $100$ random samples of a Gaussian frequency distribution for $N = 25, 50, 100, 200$. The curves are gray shaded (colored in the online version) according to their respective excess kurtosis $\gamma_2$. Note that the range of $\gamma_2$ differs considerably between the panels. The black dashed line stems from regular sampling. 
   The dotted black line marks the cut at $R_\mathrm{min}=0.2$, at which 
   the spread presented in Fig.~\ref{fig:kurt_eps} is measured. }
   \label{fig:Rmin_diff_N}
\end{figure*}

\subsection{Sample kurtosis of $g(\omega)$ determines the shape of the transition curve $R_\mathrm{min}(\epsilon)$}

 In this section, we identify the sample kurtosis as the main reason for the spread 
 of the $R_\mathrm{min}(\epsilon)$ curves, in particular for the spread of the critical values 
 of coupling at which
 $R_\mathrm{min}$ becomes nonzero.  
 Furthermore, we quantify the scaling of this spread with ensemble size. 

Figure~\ref{fig:Rmin_diff_N} presents the results of a numerical experiment,
in which we generate $100$ frequency samples of a Gaussian distribution $\mathcal{N}(0,1)$ for each of the 
ensemble sizes $N = 25, \,50,\, 100,\, 200$. For $150$ coupling strengths $\epsilon$ ranging from $0.7$ to $3.3$
and for each frequency sample, we initiate a time series from one fixed set of uniformly distributed random phases. We plot the minimum 
of $R$ after observation time $t= 10^5$ (with initial transients of length $t=10^4$)
versus coupling strength $\epsilon$. Each line in
the plot is gray shaded (colored in the online version)
according to the sample kurtosis of the respective frequency sample.

Kurtosis--the fourth standardized central moment $\beta_2 = \langle\omega_i^4\rangle\langle\omega_i^2\rangle^{-2}$--quantifies the probability weight in the tails of a probability distribution. 
The kurtosis of a Gaussian equals $3$ (\textit{mesokurtic} distribution), 
therefore comparisons among different distributions often refer to excess kurtosis $\gamma_2 = \beta_2 -3$. 
Positive excess kurtosis (\textit{leptokurtic} distribution) often indicates 
fatter tails--which in our case means many moderately extreme frequencies that require stronger coupling to join the synchronization cluster. Negative excess kurtosis (\textit{platykurtic} distribution) 
corresponds to distributions with more probability weight concentrated closely
around the mean \footnote{The authors of \cite{kurtosisBalanda1988} refine this simplified description to 
``kurtosis vaguely [is the] location- and scale-free movement of probability mass from the shoulders of a distribution 
into its center and tail''.}--here we have a broader range of small frequencies with an almost constant probability and few but extreme 
outliers \footnote{Generating a sample from a 
Gaussian with two or more clearly distinct modes (maxima) is rather improbable, we do not discuss them here. 
We just notice that
already bimodal frequency distributions exhibit a rich bifurcation 
map~\cite{PhysRevE.79.026204, PhysRevE.80.046215}.}.  
This can be understood as follows: Under the restriction of unit standard 
deviation, the few outliers in platikurtic samples must be significantly 
larger in their absolute value than the many outliers in the fat tails of 
leptokurtic samples, because they must compensate for the tightly packed 
frequencies gathered around the mean to give the same standard deviation.

\begin{figure}
% \centering
 \includegraphics[width = \columnwidth]{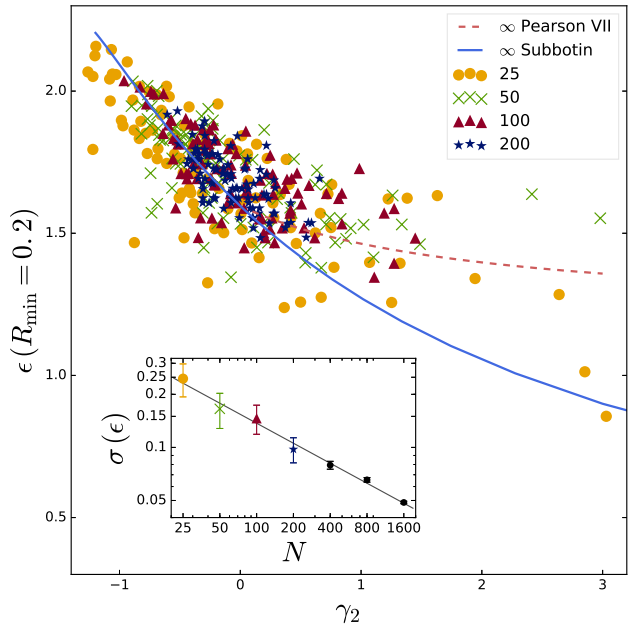}
 \caption{(color online) Coupling strength at the level $R_\mathrm{min}= 0.2$ vs. sample 
 kurtosis of the respective $g(\omega)$-sample; the same data as in Fig. \ref{fig:Rmin_diff_N} are used. 
Solid and dashed lines give predictions from numerical integration of Eq.~\eqref{eq:kura3} for two families of 
distributions with kurtosis as a parameter (cf. Sec.~\ref{kurt_infty}). Inset: 
Scaling of the standard deviation of $\epsilon$ corresponding to 
$R_\mathrm{min}=0.2$ with the number of oscillators $N$; linearly fitted by power law 
$\sigma[\epsilon(R_\mathrm{min} = 0.2)]\approx N^{-0.38}$. }\label{fig:kurt_eps}
\end{figure}

Figure~\ref{fig:Rmin_diff_N} shows that indeed the course and position of the curve $R_\mathrm{min}$ vs. $\epsilon$
is highly correlated with the sample kurtosis. One can differentiate three groups according to the
darkness (color) of the curves:
 light-gray (pink) for the most negative observed excess kurtosis, gray (green) for samples with 
 nearly vanishing $\gamma_2$, and black for the largest observed excess kurtosis. 
 These three groups are well distinguishable for all ensemble sizes, and are characterized by the following properties:
\begin{enumerate}
\item Platykurtic samples of $g(\omega)$ (light gray, negative excess kurtosis) 
lift off from $R_\mathrm{min}=0$ only at strong critical couplings. 
The curve climbs rapidly or even jumps to significant 
values of $R_\mathrm{min}\approx 0.4\dots 0.6$. These samples achieve full frequency 
synchronization at the lowest coupling strengths compared to other samples.
\item For leptokurtic samples (black, positive excess kurtosis), in contrast,  comparably weak values of coupling 
suffice to synchronize a considerable central cluster. Thereby, $R_\mathrm{min}$ grows sedately 
from a rather small $\epsilon_c^\mathrm{min}$. Due to the constraint of unit variance, 
the outlier frequencies must be quite extreme and thus are eventually synchronized only by much stronger coupling. Consequently, these samples require stronger coupling to achieve full frequency locking. 
\item Mesokurtic samples (small kurtosis, gray) naturally lie between these two extremes: they demonstrate
roughly a ``standard'' transition in the curve $R_\mathrm{min}(\epsilon)$, similar to that demonstrated by the regular sample generated by virtue of quantiles (black dashed curves).
\end{enumerate}

%Note that these are rather loose rules, as for sample sizes $\sim 100$, the contributions from higher moments are not negligible. In Sect.~\ref{kurt_infty}, we single out the role of kurtosis by integrating the infinite model.

 In Fig.~\ref{fig:kurt_eps}, we quantify some of the qualitative observations above. Here
we show the dependence of the value of the coupling parameter $\epsilon$ at which $R_\mathrm{min}$ first crosses 
the threshold $R_\mathrm{min} = 0.2$ on the excess kurtosis for different system sizes $N$. The data confirm 
 the inverse proportionality mentioned above.
 As a theoretical substantiation, we compare the results from these numerical experiments
 in finite ensembles with the kurtosis dependence of $\epsilon(R_\mathrm{min}=0.2$) in the thermodynamic  limit; see
 Sec.~\ref{sec:samp_dist} for details. The inset in Fig.~\ref{fig:kurt_eps} depicts the scaling of the standard deviation of $\epsilon$ at $R=0.2$ with system size $N$. As expected, the variability decreases in the limit $N\to\infty$ -- approximately with $\Delta\epsilon \sim N^{-0.38}$.
 
%  \begin{comment}\begin{figure}
%  \centering
%  \includegraphics[width = \columnwidth]{img/std_vs_N_tmp.pdf}
%  \caption{Scaling of standard deviation of the $\epsilon$ corresponding to 
% $R_\mathrm{min}=0.2$ with $N$. }\label{fig:std_vs_N}
% \end{figure}
% \end{comment}
% 
% 

%\input{drift_skew.tex}
\subsection{Sample skewness of $g(\omega)$ determines the drift of the global phase}\label{sec:drift}

  \begin{figure}[!htb]
 \centering\includegraphics[width = \columnwidth]{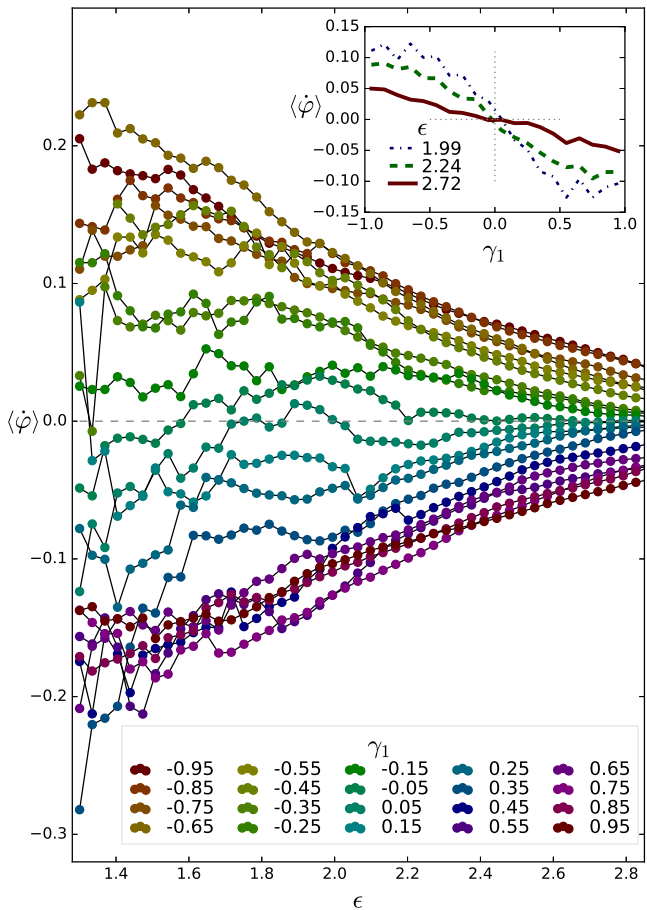}
 \caption{(color online)  Mean frequency of the macroscopic phase 
 $\Omega=\left<\dot\varphi\right>$ for $N=50$ oscillators versus coupling strength $\epsilon$. Different gray scales (colors 
 in online version) indicate different sample 
 skewnesses from top,  $\gamma_1 = -0.95$, to bottom, $\gamma_1= 0.95$. Each chain of dots represents the mean over up to $20$ phase velocities of distribution samples 
 with the same ($\pm10^{-5}$) skewness (the relative number of samples 
 with $R_\mathrm{min}>0.1$, i.e., for which a collective mode is actually defined at a given $\epsilon$, increases with coupling strength). The frequencies are calculated by averaging over the time 
 interval $10^3$.
 The inset shows cuts through the main picture at three different 
 values of $\epsilon$: phase velocity vs. skewness.}\label{fig:grad_over_skew}
\end{figure}

Random sampling of $g(\omega)$ results not only in variations of the shape of the distribution, characterized above
by kurtosis but also in deviations from symmetry with respect to the mean value (the underlying Gaussian distribution is symmetric). These can be
characterized by sample skewness $\gamma_1 = \langle\omega_i^3\rangle\langle\omega_i^2\rangle^{-3/2}$.
The main effect of skewness is a finite macroscopic frequency, $\Omega=\langle\dot\varphi\rangle_t$.  
Above, when discussing the theory in the thermodynamic limit, we argued that the macroscopic frequency
vanishes for symmetric distributions. Figure~\ref{fig:cmplx_order_time}(b) confirms that 
regular sampling via quantiles generates perfectly symmetric sets (for all $i$ there exists exactly one $j$ such that $\omega_i = -\omega_j$) and the global phase truly converges in the supercritical regime.

In Fig.~\ref{fig:grad_over_skew}, we present a numerical study on the dependence of the mean global frequency $\Omega$ 
on the skewness of $g(\omega)$ samples and on the coupling strength. First, we generate many Gaussian frequency samples and pick those with skewness close ($\pm10^{-5}$) to 1 of $20$ target values of skewness, until having $20$ samples for each. Each dot in the figure corresponds to the mean over the global phase velocities of the $20$ samples with the same skewness. As explained above, $\Omega$ is only 
meaningfully defined for finite $R_\mathrm{min}$, thus cases with $R_\mathrm{min}<0.1$ were rejected and
the curves start from finite values of $\epsilon$. The individual global phases velocities are the slopes of linear fits of the global phase after cutting a transient of $10^2$ time steps.

\begin{figure}[!htb]
 \includegraphics[width=\columnwidth]{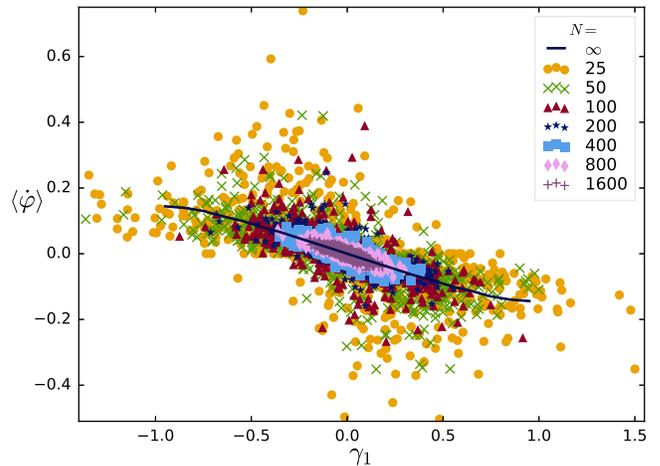}
 \caption{(color online) Average frequencies of the  mean field vs. skewness of the frequency sample for different ensemble sizes. The 
 continuous curve stems from the numerical integration of self-consistency [Eq.~\eqref{eq:kura3}] for skew-normal distributions with 
 different skewness parameters $\alpha$. Each marker shape corresponds to one ensemble size, for each of
 which $1000$ samples of a Gaussian with mean zero and variance one are generated randomly. For each sample, the time evolution of 
 $\varphi$ at $\epsilon = 2$, performed over $t=10^3$ plus $10^3$ transient, gives $\langle\dot\varphi\rangle$ represented by one 
 point in the plot.}
 \label{fig:omg_vs_skew}
\end{figure}

To clarify the dependence of the observed macroscopic frequency on the ensemble size, we perform calculations 
similar to those presented in Fig.~\ref{fig:grad_over_skew} for different values of $N$.
Figure~\ref{fig:omg_vs_skew} displays how the correlation of the global angular velocity with sample skewness 
scales with ensemble size at a fixed coupling strength $\epsilon = 2$. 
We generate $1000$ samples per $N\in[25, 50, 100, 200, 400, 800, 1600]$. In contrast to the former experiment, 
samples with all values of skewness enter the simulations.
For small sample sizes, the spread in $\Omega$ for a fixed skewness is maximal. 
With increasing $N$, samples with high skewness get less likely. 
For $N= 800, 1600$, the form of dependence of $\Omega$ vs. $\gamma_1$ coincides very well with the 
theoretical curve obtained in the thermodynamic limit
for skew-normal distributions with different $\gamma_1$, which we derive in Sec.~\ref{sec:samp_dist}.

\section{Effects of kurtosis and skewness in the thermodynamic limit}\label{sec:samp_dist}

In Sec.~\ref{sec:order_kurt}, we demonstrated that a significant part of 
the variability of the parameter $R_{\mathrm{min}}$ and of the frequency of the collective mode
$\Omega$ results from the sample variability of kurtosis and skewness, respectively.
We further support these two findings  by complementing them with calculations in the
thermodynamic limit. In infinite ensembles, 
a full analysis of stationary solutions can be performed on the basis of Eq.~\eqref{eq:kura3},
as explained in Sec.~\ref{sec:kurt} above. Instead of using a symmetric standard
distribution of frequencies, e.g., a Gaussian, we explore distributions with kurtosis and skewness
as explicit parameters. We thereby also embed findings of Refs.~\cite{Petkoski2013, Pazo2005} for rather artificial distributions into 
the context of finite-size effects.

\subsection{Effect of kurtosis}\label{kurt_infty}

We start with kurtosis. One popular family of distributions, where kurtosis is a parameter, is the \textit{Pearson type \rom{7} family} \cite{Pearson429}, but here excess kurtosis $\gamma_2$ ranges only 
from  zero to infinity. More convenient for our purposes 
is the \textit{Subbotin family}~\cite{Sub23} (sometimes also called exponential power distribution), 
with excess kurtosis ranging from $-1.2$ to $\infty$. This family covers  such 
important cases as uniform, Laplace, and Gaussian distributions. 

The Subbotin family's probability density function depends on a main 
parameter $p$ (and on an auxiliary
quantity $\sigma_p$ that serves for setting variance $\sigma$ to 1):
\begin{equation}\label{eq:Subbotin}
g(\omega; p) = \left[2\,\sigma_p\, p^{1/p}\,\Gamma(1+1/p)\right]^{-1}\,\cdot\,\exp\left(-\frac{|\omega|^p}{p\,\sigma_p^p}\right),
\end{equation}
where $\Gamma(x)$ denotes the Gamma function.
Notice the symmetry of the distribution for any $p$. The shape parameter $p$ acts inversely to the excess 
kurtosis $\gamma_2$; the exact relation is  
$\gamma_2 = \Gamma\left(1/p\right)\Gamma\left(5/p\right)\big/\left[\Gamma\left(3/p\right)\right)]^2$. 
The limit $p\to 0$, where  $\gamma_2\to\infty$, corresponds to a delta distribution.  
The Laplacian distribution has $p=1$ and $\gamma_2 = 3$; it has a peak at zero and fat tails, comparable to a Lorentz distribution.
With $p=2$, we have a Gaussian distribution with vanishing excess kurtosis. Finally, 
the case $p\to\infty$ yields a uniform distribution with excess kurtosis $\gamma_2=-1.2$.
 
The dependencies $R(\epsilon)$ for different values of the parameter $p$ can be obtained from
Eq.~\eqref{eq:kura3}. Because the imaginary part of the integral vanishes due to symmetry, we only have to calculate
the first real integral in Eq.~\eqref{eq:kura3}, which reduces to
\begin{equation}
  R= \int\limits_{-a}^a\,g(\omega;p)\,
  \sqrt{1-\left(\omega/a\right)^2}\,\mathrm{d}\omega\;,\qquad  \epsilon = a/R\;,
  \label{eq:kurtint}
 \end{equation}
 by transforming to $\omega = a\,\sin\theta$.
This expression clearly shows the role of the distribution of the probability mass to either the center or the tails of the distribution $g(\omega;p)$: The 
tails beyond $|\omega|=a$ do not contribute to the integral.

\begin{figure}[h!]
   \includegraphics[width =\columnwidth]{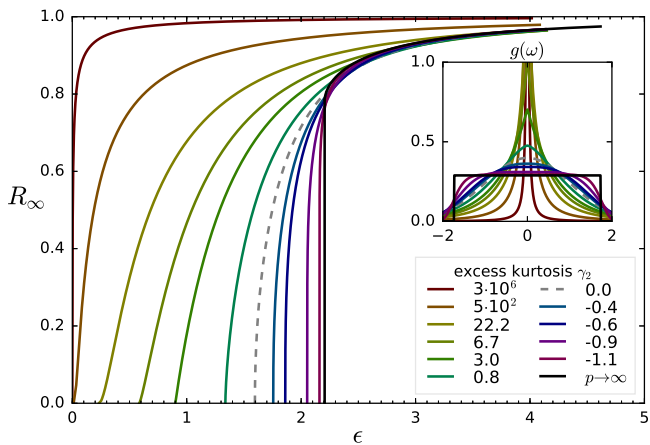}
  \caption{(color online) Kurtosis dependence of $R(\epsilon)$:
  We numerically integrate the self-consistent Eq.~(\ref{eq:kurtint}) for the Subbotin family of frequency distributions, 
  see Eq.~\eqref{eq:Subbotin}. The integral for the uniform distribution stems from the analytical solution, see Eq.~(\ref{eq:sce_uniform}) with $\omega_\mathrm{max}=\sqrt{3}$.
 The curves (from left to right)
 correspond to values of the parameter 
 $p = 0.1,\, 0.25,\, 0.5,\, 0.75,\, 1,\, 1.5,\, 2,\, 2.5,\,3,\, 5,\,10$ and to the uniform distribution with $p\to\infty$ (or likewise to the excess kurtosis values as given in the legend). The inset shows the respective probability densities.}
  \label{fig:self_cons_kurt}%\label{fig:self_cons_kurt_and_grad_over_skew_infty}
\end{figure}

We discussed the analytical integration of some of representatives $g(\omega; p)$ in the end of Sec. \ref{sec:kurt}.
For all other values of $p$, we solve the integral 
in Eq.~\eqref{eq:kurtint} numerically.
Figure~\ref{fig:self_cons_kurt} shows numerical solutions for different $p$ and thus for different excess kurtosis values $\gamma_2$, 
as well as the analytical solution for $p\to\infty$, corresponding to a uniform distribution.
The $R(\epsilon)$ curves stemming from the Subbotin family qualitatively fit
 numerical results of Fig.~\ref{fig:Rmin_diff_N}. For a quantitative correspondence, it is sufficient to 
remind that the critical value of the coupling constant $\epsilon_c$ is inversely
proportional to $g(0;p)$, which in turn grows with kurtosis. Thus one obtains inverse proportionality
of the critical coupling on kurtosis, as illustrated in Fig.~\ref{fig:kurt_eps} via the dashed curve for 
the Pearson VII family, and via the solid curve for the Subbotin family.

\subsection{Effect of skewness}\label{sec:skew_infty}

Next, we discuss the connection between skewness and the mean frequency 
of the macroscopic oscillations by exploring a family of skewed normal distributions
\begin{equation}\label{eq:skewed_g}
 g(\omega;\alpha)=\frac{1}{\sqrt{2\pi}\sigma_\alpha} \,\mathrm{e}^{-\omega^2/2\sigma_\alpha}\,\left[1+\mathrm{erf}\left(\frac{\alpha\omega}{\sqrt{2}\sigma_\alpha}\right)\right]
\end{equation}
where erf is the error function 
and the parameter $\alpha$ sets the 
skewness $\gamma_1 = \frac{4-\pi}{2}\,\alpha^3\cdot[\frac{\pi}{2}(1+\alpha^2)-\alpha^2]^{-3/2}$.

\begin{figure}[h!]
 \centering
 \includegraphics[width =\columnwidth]{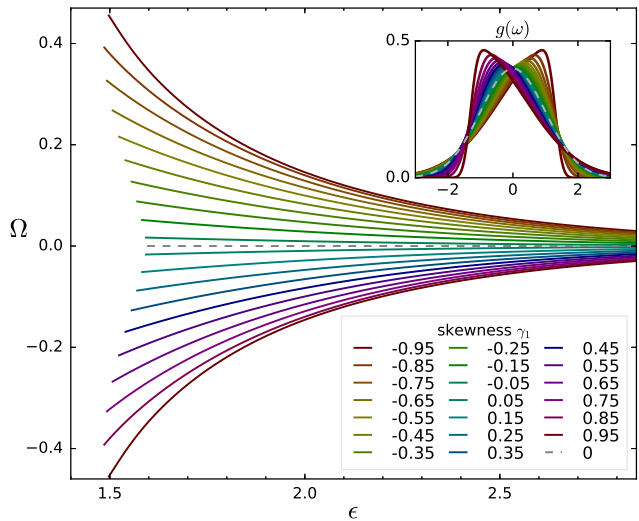}
  \caption{(color online) Frequency of the macroscopic oscillations $\Omega$ for skewed normal distributions as in Eq.~\eqref{eq:skewed_g}, obtained
  via solving the self-consistency relation Eq.~\eqref{eq:SCEimag} for 
  skewness from top $\gamma_1 = -0.95$ to bottom $\gamma_1 = 0.95$. The distributions are depicted in the inset. 
   %N
  }\label{fig:grad_over_skew_infty}
\end{figure}

Numerical solution of Eq.~\eqref{eq:kura3} for the asymmetric case requires an additional step, namely finding
the value of $\Omega_a$ for each $a$, such that the imaginary part of the 
integral in Eq.~(\ref{eq:kura3}) vanishes. We first simplify the asynchronous integral for a 
general skewed distribution ($\alpha$ fixed) to
\begin{align}\label{eq:SCEimag}
 \mathcal{I}_\mathrm{asy} &= \frac{ i}{a}\biggr[\int\limits_{a}^\infty \mathrm{d}\omega\, g(\omega+\Omega)\,\left(2\omega-\sqrt{\omega^2-a^2}\right) \nonumber\\&\quad+ \int\limits^{-a}_{-\infty }\mathrm{d}\omega\, g(\omega+\Omega)\,\left(2\omega+\sqrt{\omega^2-a^2}\right)\biggr] \overset{!}{=}0.
\end{align}
Note that this integral explicitly balances out the tails of $g(\omega)$ starting from $a$, i.e., asymmetries 
inside of $[-a,a]$ have no effect on the global phase velocity.
We iteratively approach $\Omega_a$ for each $a$ using the Newton-Raphson method. As mentioned before, 
$\Omega_a$ is unique for each $a$, but still in order not to miss any roots, we explored a range $-4<\Omega<4$ -- far 
beyond typical frequencies in $g(\omega;\alpha)$. Figure~\ref{fig:grad_over_skew_infty} shows the results;
we use the same 
skewnesses as in Fig.~\ref{fig:grad_over_skew} for a straightforward comparison. The correspondence to the numerical experiment 
with finite $N$ is apparent for large values of the coupling parameter, while close to the transition, 
where averaging in Fig.~\ref{fig:grad_over_skew} covers less than $20$ distributions, deviations are large. At a fixed value of coupling 
$\epsilon=2$, the dependence of frequency $\Omega$ on the skewness in the family Eq.~\eqref{eq:skewed_g}
is shown with a solid line in Fig.~\ref{fig:omg_vs_skew}. This dependence fits the numerical data very
well for finite large ensembles.

\section{Conclusions}
\label{sec:concl}
Motivated by the approach of Ref.~\cite{PhysRevE.85.015204}, we characterize the 
transition to synchrony in finite populations of phase oscillators as the emergence of a well-defined collective mode.
 A collective phase is well defined only if a collective amplitude never vanishes. 
Therefore, the minimum of the Kuramoto order 
parameter over a sufficiently long observation time, $R_{\text{min}}$, serves as a proper criterion for the presence of a collective mode
by ensuring a meaningful definition of the Kuramoto global phase at all times. In contrast to 
the time-averaged value of the order parameter, 
 $R_{\text{min}}$ undergoes a sharp transition. This allows us to determine a critical coupling 
strength $\epsilon_c$ for the synchronization transition in each sample. 
This transition point can be determined with high precision by extending the time series along which the minimum is retrieved. 

Furthermore, we attribute the variations of the coupling dependence of the 
explored parameter  $R_{\text{min}}$ 
to the variations of  the effective shape of the underlying frequency 
distribution, measured by kurtosis. Under the constraint of unit variance, kurtosis of a finite frequency sample 
allows us to distinguish whether either most frequencies are crowded close to the mean and few extreme outliers balance the 
variance or all frequencies are rather broadly distributed
but  without extremes. 
All samples lie between these two 
limiting cases of platykurtic and leptokurtic samples, respectively. In leptokurtic samples, 
the central frequencies synchronize at comparably low coupling strength and their critical coupling is accordingly 
small. The few extreme outliers require much stronger coupling to adjust their frequency to the central cluster. 
In contrast, platykurtic samples have comparably large values of critical coupling  $\epsilon_c$
but then rapidly climb or even jump to high values of the order parameter, eventually reaching full frequency 
locking at lower coupling strengths compared to platykurtic samples. These properties are well reproduced 
in the thermodynamic limit, shown by numerically calculating the stationary solution for the order parameter for a family of distributions with selectable kurtosis.

The second observation in finite ensembles where we were able to treat
the thermodynamic limit in a similar way is the dependence of the mean 
frequency of the global phase on the sample skewness of some natural frequency distribution. The mean frequency 
and skewness are roughly anticorrelated. 
Ensembles with perfectly symmetric frequency samples converge to a constant, nonrotating 
phase at any coupling strength above transition.

Note that the considerations regarding the effects of kurtosis and skewness on 
the dynamics in the thermodynamic limit cannot straightforwardly be extended 
to distributions that do not possess the corresponding moments. 
A prominent example due to its analytical tractability is
the Lorentz distribution. Actually, the only representative of the 
important family of
alpha-stable distributions with defined moments other than the mean is 
the Gaussian distribution. Though moments are undefined, this distribution 
family has parameters for asymmetry and shape.

The approach of Sec.~\ref{sec:kurt} for the thermodynamic limit can be 
applied to distributions of any asymmetry and shape, 
provided the integrals in Eq.~\eqref{eq:kura3} converge.
A remaining challenge lies in finding a meaningful procedure to relate sample asymmetry and sample shape for
finite samples of the Lorentz distribution to families of continuous distributions,
as we presented for the Gaussian ensemble in Sec.~\ref{sec:samp_dist}. 

Conceptually, a fat-tailed distribution of frequencies 
potentially contradicts the model assumption of nearly identical 
oscillators. In finite samples, the effect of extreme 
outliers on the collective dynamics is, however, comparably small due to a 
strong separation of time scales. The usage of the value $R_\mathrm{min}$ as an indicator for the transition is
therefore applicable to the Lorentz ensemble and to other fat-tailed distributions.

In this paper we applied the criterion of the 
existence of a global oscillatory mode,
based on the existence of the global phase,
to the Kuramoto model. It would be interesting to explore finite-size effects on collective modes
also for other types of synchronization transitions, where, e.g., all oscillators remain unlocked 
and the collective mode is related to partial synchrony, see Refs.~\cite{vanVreeswijk-96,Clusella_etal-16}.

%\color{red}write here why this would be nice.\color{black} 
% 
% \begin{comment}
% \begin{itemize}
%  \item open questions and perspectives:
%  \item: 
%  \item To what extend is the dynamics above transition really quasi-periodic with few frequencies or is there always a chaotic component present?
%  \item application to complex network structures?
%  \item .. maybe there is some theory around the SCE plus Bessel functions?
% \end{itemize}
% \end{comment}

\begin{acknowledgments}
This paper was developed within the scope of the IRTG 1740/TRP 2015/50122-0, 
funded by the DFG/ FAPESP. In studies presented in Sec.~\ref{sec:samp_dist}, 
A.P. was supported by the Russian Science Foundation (Grant No.\ 17-12-01534). 
We thank Ralf Toenjes for valuable discussions.
\end{acknowledgments}

%\bibliography{bib_all_01Nov2016.bib}

\begin{thebibliography}{43}%
\makeatletter
\providecommand \@ifxundefined [1]{%
 \@ifx{#1\undefined}
}%
\providecommand \@ifnum [1]{%
 \ifnum #1\expandafter \@firstoftwo
 \else \expandafter \@secondoftwo
 \fi
}%
\providecommand \@ifx [1]{%
 \ifx #1\expandafter \@firstoftwo
 \else \expandafter \@secondoftwo
 \fi
}%
\providecommand \natexlab [1]{#1}%
\providecommand \enquote  [1]{``#1''}%
\providecommand \bibnamefont  [1]{#1}%
\providecommand \bibfnamefont [1]{#1}%
\providecommand \citenamefont [1]{#1}%
\providecommand \href@noop [0]{\@secondoftwo}%
\providecommand \href [0]{\begingroup \@sanitize@url \@href}%
\providecommand \@href[1]{\@@startlink{#1}\@@href}%
\providecommand \@@href[1]{\endgroup#1\@@endlink}%
\providecommand \@sanitize@url [0]{\catcode `\\12\catcode `\$12\catcode
  `\&12\catcode `\#12\catcode `\^12\catcode `\_12\catcode `\%12\relax}%
\providecommand \@@startlink[1]{}%
\providecommand \@@endlink[0]{}%
\providecommand \url  [0]{\begingroup\@sanitize@url \@url }%
\providecommand \@url [1]{\endgroup\@href {#1}{\urlprefix }}%
\providecommand \urlprefix  [0]{URL }%
\providecommand \Eprint [0]{\href }%
\providecommand \doibase [0]{http://dx.doi.org/}%
\providecommand \selectlanguage [0]{\@gobble}%
\providecommand \bibinfo  [0]{\@secondoftwo}%
\providecommand \bibfield  [0]{\@secondoftwo}%
\providecommand \translation [1]{[#1]}%
\providecommand \BibitemOpen [0]{}%
\providecommand \bibitemStop [0]{}%
\providecommand \bibitemNoStop [0]{.\EOS\space}%
\providecommand \EOS [0]{\spacefactor3000\relax}%
\providecommand \BibitemShut  [1]{\csname bibitem#1\endcsname}%
\let\auto@bib@innerbib\@empty
%</preamble>
\bibitem [{\citenamefont {Nixon}\ \emph {et~al.}(2013)\citenamefont {Nixon},
  \citenamefont {Ronen}, \citenamefont {Friesem},\ and\ \citenamefont
  {Davidson}}]{Nixon_etal-13}%
  \BibitemOpen
  \bibfield  {author} {\bibinfo {author} {\bibfnamefont {M.}~\bibnamefont
  {Nixon}}, \bibinfo {author} {\bibfnamefont {E.}~\bibnamefont {Ronen}},
  \bibinfo {author} {\bibfnamefont {A.~A.}\ \bibnamefont {Friesem}}, \ and\
  \bibinfo {author} {\bibfnamefont {N.}~\bibnamefont {Davidson}},\ }\href
  {\doibase 10.1103/PhysRevLett.110.184102} {\bibfield  {journal} {\bibinfo
  {journal} {Phys. Rev. Lett.}\ }\textbf {\bibinfo {volume} {110}},\ \bibinfo
  {pages} {184102} (\bibinfo {year} {2013})}\BibitemShut {NoStop}%
\bibitem [{\citenamefont {Kiss}\ \emph {et~al.}(2002)\citenamefont {Kiss},
  \citenamefont {Zhai},\ and\ \citenamefont {Hudson}}]{Kiss2002}%
  \BibitemOpen
  \bibfield  {author} {\bibinfo {author} {\bibfnamefont {I.~Z.}\ \bibnamefont
  {Kiss}}, \bibinfo {author} {\bibfnamefont {Y.}~\bibnamefont {Zhai}}, \ and\
  \bibinfo {author} {\bibfnamefont {J.~L.}\ \bibnamefont {Hudson}},\ }\href
  {\doibase 10.1126/science.1070757} {\bibfield  {journal} {\bibinfo  {journal}
  {Science (N.Y.)}\ }\textbf {\bibinfo {volume} {296}},\ \bibinfo
  {pages} {1676} (\bibinfo {year} {2002})}\BibitemShut {NoStop}%
\bibitem [{\citenamefont {Motter}\ \emph {et~al.}(2013)\citenamefont {Motter},
  \citenamefont {Myers}, \citenamefont {Anghel},\ and\ \citenamefont
  {Nishikawa}}]{Motter-13}%
  \BibitemOpen
  \bibfield  {author} {\bibinfo {author} {\bibfnamefont {A.~E.}\ \bibnamefont
  {Motter}}, \bibinfo {author} {\bibfnamefont {S.~A.}\ \bibnamefont {Myers}},
  \bibinfo {author} {\bibfnamefont {M.}~\bibnamefont {Anghel}}, \ and\ \bibinfo
  {author} {\bibfnamefont {T.}~\bibnamefont {Nishikawa}},\ }\href {\doibase
  doi:10.1038/nphys2535} {\bibfield  {journal} {\bibinfo  {journal} {Nat.
  Phys.}\ }\textbf {\bibinfo {volume} {9}},\ \bibinfo {pages} {191} (\bibinfo
  {year} {2013})}\BibitemShut {NoStop}%
\bibitem [{\citenamefont {Antzoulatos}\ and\ \citenamefont
  {Miller}()}]{Miller2014}%
  \BibitemOpen
  \bibfield  {author} {\bibinfo {author} {\bibfnamefont {E.~G.}\ \bibnamefont
  {Antzoulatos}}\ and\ \bibinfo {author} {\bibfnamefont {E.~K.}\ \bibnamefont
  {Miller}},\ }\href {\doibase 10.1016/j.neuron.2014.05.005} {\bibfield
  {journal} {\bibinfo  {journal} {Neuron}\ }\textbf {\bibinfo {volume} {83}},\
  \bibinfo {pages} {216}}\BibitemShut {NoStop}%
\bibitem [{\citenamefont {Winfree}(1980)}]{Winfree1980}%
  \BibitemOpen
  \bibfield  {author} {\bibinfo {author} {\bibfnamefont {A.~T.}\ \bibnamefont
  {Winfree}},\ }\href {\doibase 10.1007/978-1-4757-3484-3} {\emph {\bibinfo
  {title} {{The Geometry of Biological Time}}}},\ \bibinfo {edition} {1st}\
  ed.\ (\bibinfo  {publisher} {Springer, Berlin,},\ \bibinfo {year}
  {1980})\BibitemShut {NoStop}%
\bibitem [{\citenamefont {Kuramoto}(1975)}]{Kuramoto1975}%
  \BibitemOpen
  \bibfield  {author} {\bibinfo {author} {\bibfnamefont {Y.}~\bibnamefont
  {Kuramoto}},\ }\href {\doibase 10.1007/BFb0013365} {\bibfield  {journal}
  {\bibinfo  {journal} {Math. Probl. Theor. Phys.}\ }\textbf
  {\bibinfo {volume} {39}},\ \bibinfo {pages} {420} (\bibinfo {year}
  {1975})}\BibitemShut {NoStop}%
\bibitem [{\citenamefont {Kuramoto}(1984)}]{Kuramoto1984}%
  \BibitemOpen
  \bibfield  {author} {\bibinfo {author} {\bibfnamefont {Y.}~\bibnamefont
  {Kuramoto}},\ }\href {\doibase 10.1007/978-3-642-69689-3} {\emph {\bibinfo
  {title} {{Chemical Oscillations, Waves, and Turbulence}}}},\ \bibinfo
  {series} {Springer Series in Synergetics}, Vol.~\bibinfo {volume} {19}\
  (\bibinfo  {publisher} {Springer, Berlin},\ \bibinfo {year}
  {1984})\BibitemShut {NoStop}%
\bibitem [{\citenamefont {Daido}(1996)}]{Daido-96a}%
  \BibitemOpen
  \bibfield  {author} {\bibinfo {author} {\bibfnamefont {H.}~\bibnamefont
  {Daido}},\ }\href {\doibase 10.1103/PhysRevLett.77.1406} {\bibfield
  {journal} {\bibinfo  {journal} {Phys. Rev. Lett.}\ }\textbf {\bibinfo
  {volume} {77}},\ \bibinfo {pages} {1406} (\bibinfo {year}
  {1996})}\BibitemShut {NoStop}%
\bibitem [{\citenamefont {Komarov}\ and\ \citenamefont
  {Pikovsky}(2014)}]{Komarov2014}%
  \BibitemOpen
  \bibfield  {author} {\bibinfo {author} {\bibfnamefont {M.}~\bibnamefont
  {Komarov}}\ and\ \bibinfo {author} {\bibfnamefont {A.}~\bibnamefont
  {Pikovsky}},\ }\href {\doibase 10.1016/j.physd.2014.09.002} {\bibfield
  {journal} {\bibinfo  {journal} {Physica D: Nonlinear Phenomena}\ }\textbf
  {\bibinfo {volume} {289}},\ \bibinfo {pages} {18} (\bibinfo {year} {2014})}
  \BibitemShut
  {NoStop}%
\bibitem [{\citenamefont {Sakaguchi}\ and\ \citenamefont
  {Kuramoto}(1986)}]{Sakaguchi_Kuramoto86}%
  \BibitemOpen
  \bibfield  {author} {\bibinfo {author} {\bibfnamefont {H.}~\bibnamefont
  {Sakaguchi}}\ and\ \bibinfo {author} {\bibfnamefont {Y.}~\bibnamefont
  {Kuramoto}},\ }\href {\doibase 10.1143/PTP.76.576} {\bibfield  {journal}
  {\bibinfo  {journal} {Progress of Theoretical Physics}\ }\textbf {\bibinfo
  {volume} {76}},\ \bibinfo {pages} {576} (\bibinfo {year} {1986})}\BibitemShut
  {NoStop}%
\bibitem [{\citenamefont {Acebr{\'{o}}n}\ \emph {et~al.}(2005)\citenamefont
  {Acebr{\'{o}}n}, \citenamefont {Bonilla}, \citenamefont {Vicente},
  \citenamefont {Ritort},\ and\ \citenamefont {Spigler}}]{Acebron2005}%
  \BibitemOpen
  \bibfield  {author} {\bibinfo {author} {\bibfnamefont {J.~A.}\ \bibnamefont
  {Acebr{\'{o}}n}}, \bibinfo {author} {\bibfnamefont {L.~L.}\ \bibnamefont
  {Bonilla}}, \bibinfo {author} {\bibfnamefont {C.~J.~P.}\ \bibnamefont
  {Vicente}}, \bibinfo {author} {\bibfnamefont {F.}~\bibnamefont {Ritort}}, \
  and\ \bibinfo {author} {\bibfnamefont {R.}~\bibnamefont {Spigler}},\ }\href
  {\doibase 10.1103/RevModPhys.77.137} {\bibfield  {journal} {\bibinfo
  {journal} {Reviews of Modern Physics}\ }\textbf {\bibinfo {volume} {77}},\
  \bibinfo {pages} {137} (\bibinfo {year} {2005})} \BibitemShut {NoStop}%
\bibitem [{\citenamefont {Arenas}\ \emph {et~al.}(2008)\citenamefont {Arenas},
  \citenamefont {Kurths}, \citenamefont {Moreno}, \citenamefont
  {D{\'{i}}az-Guilera},\ and\ \citenamefont {Zhou}}]{Arenas2008}%
  \BibitemOpen
  \bibfield  {author} {\bibinfo {author} {\bibfnamefont {A.}~\bibnamefont
  {Arenas}}, \bibinfo {author} {\bibfnamefont {J.}~\bibnamefont {Kurths}},
  \bibinfo {author} {\bibfnamefont {Y.}~\bibnamefont {Moreno}}, \bibinfo
  {author} {\bibfnamefont {A.}~\bibnamefont {D{\'{i}}az-Guilera}}, \ and\
  \bibinfo {author} {\bibfnamefont {C.}~\bibnamefont {Zhou}},\ }\href {\doibase
  10.1016/j.physrep.2008.09.002} {\bibfield  {journal} {\bibinfo  {journal}
  {Physics Reports}\ }\textbf {\bibinfo {volume} {469}},\ \bibinfo {pages} {1}
  (\bibinfo {year} {2008})} \BibitemShut {NoStop}%
\bibitem [{\citenamefont {Watanabe}\ and\ \citenamefont
  {Strogatz}(1994)}]{Watanabe1994}%
  \BibitemOpen
  \bibfield  {author} {\bibinfo {author} {\bibfnamefont {S.}~\bibnamefont
  {Watanabe}}\ and\ \bibinfo {author} {\bibfnamefont {S.~H.}\ \bibnamefont
  {Strogatz}},\ }\href {\doibase 10.1016/0167-2789(94)90196-1} {\bibfield
  {journal} {\bibinfo  {journal} {Physica D: Nonlinear Phenomena}\ }\textbf
  {\bibinfo {volume} {74}},\ \bibinfo {pages} {197} (\bibinfo {year}
  {1994})}\BibitemShut {NoStop}%
\bibitem [{\citenamefont {Ott}\ and\ \citenamefont {Antonsen}(2008)}]{Ott2008}%
  \BibitemOpen
  \bibfield  {author} {\bibinfo {author} {\bibfnamefont {E.}~\bibnamefont
  {Ott}}\ and\ \bibinfo {author} {\bibfnamefont {T.~M.}\ \bibnamefont
  {Antonsen}},\ }\href {\doibase 10.1063/1.2930766} {\bibfield  {journal}
  {\bibinfo  {journal} {Chaos}\ }\textbf {\bibinfo {volume} {18}} (\bibinfo
  {year} {2008}),\ 10.1063/1.2930766}\href{\doibase 10.1016/0167-2789(94)90196-1}\BibitemShut {NoStop}%
\bibitem [{\citenamefont {Prindle}\ \emph {et~al.}(2012)\citenamefont
  {Prindle}, \citenamefont {Samayoa}, \citenamefont {Razinkov}, \citenamefont
  {Danino}, \citenamefont {Tsimring},\ and\ \citenamefont
  {Hasty}}]{Prindle_etal-12}%
  \BibitemOpen
  \bibfield  {author} {\bibinfo {author} {\bibfnamefont {A.}~\bibnamefont
  {Prindle}}, \bibinfo {author} {\bibfnamefont {P.}~\bibnamefont {Samayoa}},
  \bibinfo {author} {\bibfnamefont {I.}~\bibnamefont {Razinkov}}, \bibinfo
  {author} {\bibfnamefont {T.}~\bibnamefont {Danino}}, \bibinfo {author}
  {\bibfnamefont {L.~S.}\ \bibnamefont {Tsimring}}, \ and\ \bibinfo {author}
  {\bibfnamefont {J.}~\bibnamefont {Hasty}},\ }\href {\doibase
  10.1038/nature10722} {\bibfield  {journal} {\bibinfo  {journal} {Nature}\
  }\textbf {\bibinfo {volume} {481}},\ \bibinfo {pages} {39} (\bibinfo {year}
  {2012})}\BibitemShut {NoStop}%
\bibitem [{\citenamefont {Weber}\ \emph {et~al.}(2012)\citenamefont {Weber},
  \citenamefont {Prokazov}, \citenamefont {Zuschratter},\ and\ \citenamefont
  {Hauser}}]{Hauser2012}%
  \BibitemOpen
  \bibfield  {author} {\bibinfo {author} {\bibfnamefont {A.}~\bibnamefont
  {Weber}}, \bibinfo {author} {\bibfnamefont {Y.}~\bibnamefont {Prokazov}},
  \bibinfo {author} {\bibfnamefont {W.}~\bibnamefont {Zuschratter}}, \ and\
  \bibinfo {author} {\bibfnamefont {M.~J.~B.}\ \bibnamefont {Hauser}},\ }\href
  {\doibase 10.1371/journal.pone.0043276} {\bibfield  {journal} {\bibinfo
  {journal} {PLoS ONE}\ }\textbf {\bibinfo {volume} {7}},\ \bibinfo {pages} {1}
  (\bibinfo {year} {2012})}\BibitemShut {NoStop}%
\bibitem [{\citenamefont {Temirbayev}\ \emph {et~al.}(2012)\citenamefont
  {Temirbayev}, \citenamefont {Zhanabaev}, \citenamefont {Tarasov},
  \citenamefont {Ponomarenko},\ and\ \citenamefont
  {Rosenblum}}]{PhysRevE.85.015204}%
  \BibitemOpen
  \bibfield  {author} {\bibinfo {author} {\bibfnamefont {A.~A.}\ \bibnamefont
  {Temirbayev}}, \bibinfo {author} {\bibfnamefont {Z.~Z.}\ \bibnamefont
  {Zhanabaev}}, \bibinfo {author} {\bibfnamefont {S.~B.}\ \bibnamefont
  {Tarasov}}, \bibinfo {author} {\bibfnamefont {V.~I.}\ \bibnamefont
  {Ponomarenko}}, \ and\ \bibinfo {author} {\bibfnamefont {M.}~\bibnamefont
  {Rosenblum}},\ }\href {\doibase 10.1103/PhysRevE.85.015204} {\bibfield
  {journal} {\bibinfo  {journal} {Phys. Rev. E}\ }\textbf {\bibinfo {volume}
  {85}},\ \bibinfo {pages} {015204} (\bibinfo {year} {2012})}\BibitemShut
  {NoStop}%
\bibitem [{\citenamefont {Popovych}\ \emph {et~al.}(2005)\citenamefont
  {Popovych}, \citenamefont {Maistrenko},\ and\ \citenamefont
  {Tass}}]{Popovych-Maistrenko-Tass-05}%
  \BibitemOpen
  \bibfield  {author} {\bibinfo {author} {\bibfnamefont {O.~V.}\ \bibnamefont
  {Popovych}}, \bibinfo {author} {\bibfnamefont {Y.~L.}\ \bibnamefont
  {Maistrenko}}, \ and\ \bibinfo {author} {\bibfnamefont {P.~A.}\ \bibnamefont
  {Tass}},\ }\href {\doibase 10.1103/PhysRevE.71.065201} {\bibfield  {journal}
  {\bibinfo  {journal} {Phys. Rev. E}\ }\textbf {\bibinfo {volume} {71}},\
  \bibinfo {pages} {065201} (\bibinfo {year} {2005})}\BibitemShut {NoStop}%
\bibitem [{\citenamefont {Daido}(1987)}]{Daido1987}%
  \BibitemOpen
  \bibfield  {author} {\bibinfo {author} {\bibfnamefont {H.}~\bibnamefont
  {Daido}},\ }\href {\doibase 10.1088/0305-4470/20/10/002} {\bibfield
  {journal} {\bibinfo  {journal} {Journal of Physics A: Mathematical and
  General}\ }\textbf {\bibinfo {volume} {20}},\ \bibinfo {pages} {L629}
  (\bibinfo {year} {1987})}\BibitemShut {NoStop}%
\bibitem [{\citenamefont {Daido}(1990)}]{Daido1990}%
  \BibitemOpen
  \bibfield  {author} {\bibinfo {author} {\bibfnamefont {H.}~\bibnamefont
  {Daido}},\ }\href {\doibase 10.1007/BF01025993} {\bibfield  {journal}
  {\bibinfo  {journal} {Journal of Statistical Physics}\ }\textbf {\bibinfo
  {volume} {60}},\ \bibinfo {pages} {753} (\bibinfo {year} {1990})}\BibitemShut
  {NoStop}%
\bibitem [{\citenamefont {Son}\ and\ \citenamefont {Hong}(2010)}]{Son2010}%
  \BibitemOpen
  \bibfield  {author} {\bibinfo {author} {\bibfnamefont {S.~W.}\ \bibnamefont
  {Son}}\ and\ \bibinfo {author} {\bibfnamefont {H.}~\bibnamefont {Hong}},\
  }\href {\doibase 10.1103/PhysRevE.81.061125} {\bibfield  {journal} {\bibinfo
  {journal} {Physical Review E - Statistical, Nonlinear, and Soft Matter
  Physics}\ }\textbf {\bibinfo {volume} {81}},\ \bibinfo {pages} {1} (\bibinfo
  {year} {2010})}\BibitemShut {NoStop}%
\bibitem [{\citenamefont {Hong}\ \emph
  {et~al.}(2007{\natexlab{a}})\citenamefont {Hong}, \citenamefont
  {Chat{\'{e}}}, \citenamefont {Park},\ and\ \citenamefont {Tang}}]{Hong2007}%
  \BibitemOpen
  \bibfield  {author} {\bibinfo {author} {\bibfnamefont {H.}~\bibnamefont
  {Hong}}, \bibinfo {author} {\bibfnamefont {H.}~\bibnamefont {Chat{\'{e}}}},
  \bibinfo {author} {\bibfnamefont {H.}~\bibnamefont {Park}}, \ and\ \bibinfo
  {author} {\bibfnamefont {L.~H.}\ \bibnamefont {Tang}},\ }\href {\doibase
  10.1103/PhysRevLett.99.184101} {\bibfield  {journal} {\bibinfo  {journal}
  {Phys. Rev. Lett}\ }\textbf {\bibinfo {volume} {99}},\ \bibinfo
  {pages} {1} (\bibinfo {year} {2007}{\natexlab{a}})},\ \Eprint
  {http://arxiv.org/abs/0701646} {arXiv:0701646} \BibitemShut {NoStop}%
\bibitem [{\citenamefont {Hong}\ \emph {et~al.}(2015)\citenamefont {Hong},
  \citenamefont {Chat{\'{e}}}, \citenamefont {Tang},\ and\ \citenamefont
  {Park}}]{Hong2015}%
  \BibitemOpen
  \bibfield  {author} {\bibinfo {author} {\bibfnamefont {H.}~\bibnamefont
  {Hong}}, \bibinfo {author} {\bibfnamefont {H.}~\bibnamefont {Chat{\'{e}}}},
  \bibinfo {author} {\bibfnamefont {L.~H.}\ \bibnamefont {Tang}}, \ and\
  \bibinfo {author} {\bibfnamefont {H.}~\bibnamefont {Park}},\ }\href {\doibase
  10.1103/PhysRevE.92.022122} {\bibfield  {journal} {\bibinfo  {journal}
  {Physical Review E - Statistical, Nonlinear, and Soft Matter Physics}\
  }\textbf {\bibinfo {volume} {92}},\ \bibinfo {pages} {1} (\bibinfo {year}
  {2015})}
  \BibitemShut {NoStop}%
\bibitem [{\citenamefont {Basnarkov}\ and\ \citenamefont
  {Urumov}(2007)}]{Basnarkov2007}%
  \BibitemOpen
  \bibfield  {author} {\bibinfo {author} {\bibfnamefont {L.}~\bibnamefont
  {Basnarkov}}\ and\ \bibinfo {author} {\bibfnamefont {V.}~\bibnamefont
  {Urumov}},\ }\href {\doibase 10.1103/PhysRevE.76.057201} {\bibfield
  {journal} {\bibinfo  {journal} {Physical Review E - Statistical, Nonlinear,
  and Soft Matter Physics}\ }\textbf {\bibinfo {volume} {76}},\ \bibinfo
  {pages} {1} (\bibinfo {year} {2007})}\BibitemShut {NoStop}%
\bibitem [{\citenamefont {Paz{\'{o}}}(2005)}]{Pazo2005}%
  \BibitemOpen
  \bibfield  {author} {\bibinfo {author} {\bibfnamefont {D.}~\bibnamefont
  {Paz{\'{o}}}},\ }\href {\doibase 10.1103/PhysRevE.72.046211} {\bibfield
  {journal} {\bibinfo  {journal} {Physical Review E - Statistical, Nonlinear,
  and Soft Matter Physics}\ }\textbf {\bibinfo {volume} {72}},\ \bibinfo
  {pages} {1} (\bibinfo {year} {2005})} \BibitemShut {NoStop}%
\bibitem [{\citenamefont {Paz\'o}\ and\ \citenamefont
  {Montbri\'o}(2009)}]{PhysRevE.80.046215}%
  \BibitemOpen
  \bibfield  {author} {\bibinfo {author} {\bibfnamefont {D.}~\bibnamefont
  {Paz\'o}}\ and\ \bibinfo {author} {\bibfnamefont {E.}~\bibnamefont
  {Montbri\'o}},\ }\href {\doibase 10.1103/PhysRevE.80.046215} {\bibfield
  {journal} {\bibinfo  {journal} {Phys. Rev. E}\ }\textbf {\bibinfo {volume}
  {80}},\ \bibinfo {pages} {046215} (\bibinfo {year} {2009})}\BibitemShut
  {NoStop}%
\bibitem [{\citenamefont {Omel'chenko}\ and\ \citenamefont
  {Wolfrum}(2012)}]{Omelchenko-Wolfrum-12}%
  \BibitemOpen
  \bibfield  {author} {\bibinfo {author} {\bibfnamefont {O.~E.}\ \bibnamefont
  {Omel'chenko}}\ and\ \bibinfo {author} {\bibfnamefont {M.}~\bibnamefont
  {Wolfrum}},\ }\href
  {https://journals.aps.org/prl/abstract/10.1103/PhysRevLett.109.164101}
  {\bibfield  {journal} {\bibinfo  {journal} {Phys. Rev. Lett.}\ }\textbf
  {\bibinfo {volume} {109}},\ \bibinfo {pages} {164101} (\bibinfo {year}
  {2012})}\BibitemShut {NoStop}%
\bibitem [{\citenamefont {Omel'chenko}\ and\ \citenamefont
  {Wolfrum}(2013)}]{Omelchenko-Wolfrum-13}%
  \BibitemOpen
  \bibfield  {author} {\bibinfo {author} {\bibfnamefont {O.~E.}\ \bibnamefont
  {Omel'chenko}}\ and\ \bibinfo {author} {\bibfnamefont {M.}~\bibnamefont
  {Wolfrum}},\ }\href
  {http://www.sciencedirect.com/science/article/pii/S016727891300239X}
  {\bibfield  {journal} {\bibinfo  {journal} {Physica D}\ }\textbf {\bibinfo
  {volume} {263}},\ \bibinfo {pages} {74} (\bibinfo {year} {2013})}\BibitemShut
  {NoStop}%
\bibitem [{\citenamefont {Zhang}\ \emph {et~al.}(2017)\citenamefont {Zhang},
  \citenamefont {Pikovsky},\ and\ \citenamefont {Liu}}]{ZhangPikovskyLiu2017}%
  \BibitemOpen
  \bibfield  {author} {\bibinfo {author} {\bibfnamefont {X.}~\bibnamefont
  {Zhang}}, \bibinfo {author} {\bibfnamefont {A.}~\bibnamefont {Pikovsky}}, \
  and\ \bibinfo {author} {\bibfnamefont {Z.}~\bibnamefont {Liu}},\ }\href
  {https://www.nature.com/articles/s41598-017-02283-1} {\bibfield  {journal}
  {\bibinfo  {journal} {Nat. Sci. Rep.}\ }\textbf {\bibinfo {volume}
  {7}} (\bibinfo {year} {2017})}\BibitemShut {NoStop}%
\bibitem [{\citenamefont {Abramowitz}\ and\ \citenamefont
  {Stegun}(1965)}]{Abramowitz}%
  \BibitemOpen
  \bibfield  {author} {\bibinfo {author} {\bibfnamefont {M.}~\bibnamefont
  {Abramowitz}}\ and\ \bibinfo {author} {\bibfnamefont {I.}~\bibnamefont
  {Stegun}},\ }\href {http://people.math.sfu.ca/~cbm/aands/} {\emph {\bibinfo
  {title} {Handbook of Mathematical Functions}}}\ (\bibinfo  {publisher} {Dover, London,},\ \bibinfo {year} {1965})\BibitemShut {NoStop}% 
\bibitem [{\citenamefont {Hong}\ \emph
  {et~al.}(2007{\natexlab{b}})\citenamefont {Hong}, \citenamefont {Park},\ and\
  \citenamefont {Tang}}]{Hong2007a}%
  \BibitemOpen
  \bibfield  {author} {\bibinfo {author} {\bibfnamefont {H.}~\bibnamefont
  {Hong}}, \bibinfo {author} {\bibfnamefont {H.}~\bibnamefont {Park}}, \ and\
  \bibinfo {author} {\bibfnamefont {L.~H.}\ \bibnamefont {Tang}},\ }\href
  {\doibase 10.1103/PhysRevE.76.066104} {\bibfield  {journal} {\bibinfo
  {journal} {Physical Review E - Statistical, Nonlinear, and Soft Matter
  Physics}\ }\textbf {\bibinfo {volume} {76}},\ \bibinfo {pages} {1} (\bibinfo
  {year} {2007}{\natexlab{b}})},\ \Eprint {http://arxiv.org/abs/0710.1137}
  {arXiv:0710.1137} \BibitemShut {NoStop}%
\bibitem [{\citenamefont {Bronski}\ \emph {et~al.}(2012)\citenamefont
  {Bronski}, \citenamefont {DeVille},\ and\ \citenamefont
  {Park}}]{Bronski2012}%
  \BibitemOpen
  \bibfield  {author} {\bibinfo {author} {\bibfnamefont {J.~C.}\ \bibnamefont
  {Bronski}}, \bibinfo {author} {\bibfnamefont {L.}~\bibnamefont {DeVille}}, \
  and\ \bibinfo {author} {\bibfnamefont {M.~J.}\ \bibnamefont {Park}},\ }\href
  {\doibase 10.1063/1.4745197} {\bibfield  {journal} {\bibinfo  {journal}
  {Chaos}\ }\textbf
  {\bibinfo {volume} {22}},\ \bibinfo {pages} {033133} (\bibinfo {year}
  {2012})}\BibitemShut {NoStop}%
\bibitem [{\citenamefont {Chiba}\ and\ \citenamefont
  {Nishikawa}(2011)}]{Chiba-Nishikawa-11}%
  \BibitemOpen
  \bibfield  {author} {\bibinfo {author} {\bibfnamefont {H.}~\bibnamefont
  {Chiba}}\ and\ \bibinfo {author} {\bibfnamefont {I.}~\bibnamefont
  {Nishikawa}},\ }\href {\doibase 10.1063/1.3647317} {\bibfield  {journal}
  {\bibinfo  {journal} {Chaos}\ }\textbf {\bibinfo {volume} {21}},\ \bibinfo
  {eid} {043103} (\bibinfo {year} {2011})}\BibitemShut {NoStop}%
\bibitem [{\citenamefont {Dietert}(2016)}]{Dietert-16}%
  \BibitemOpen
  \bibfield  {author} {\bibinfo {author} {\bibfnamefont {H.}~\bibnamefont
  {Dietert}},\ }\href {\doibase 10.1016/j.matpur.2015.11.001} {\bibfield
  {journal} {\bibinfo  {journal} {J. Math. Pur. Appl.}\ }\textbf {\bibinfo {volume} {105}},\ \bibinfo {pages} {451
  } (\bibinfo {year} {2016})}\BibitemShut {NoStop}%
\bibitem [{\citenamefont {Choi}\ \emph {et~al.}(2017)\citenamefont {Choi},
  \citenamefont {Ha},\ and\ \citenamefont {Kahng}}]{Choi2013}%
  \BibitemOpen
  \bibfield  {author} {\bibinfo {author} {\bibfnamefont {C.}~\bibnamefont
  {Choi}}, \bibinfo {author} {\bibfnamefont {M.}~\bibnamefont {Ha}}, \
  and\ \bibinfo {author} {\bibfnamefont {B.}~\bibnamefont {Kahng}},\ }\href
  {\doibase 10.1103/PhysRevE.88.032126}{\bibfield  {journal}
  {\bibinfo  {journal} {Phys. Rev E.}\ }\textbf {\bibinfo {volume}
  {88}},\ \bibinfo {pages} {032126} (\bibinfo {year} {2013})}\BibitemShut {NoStop}%
\bibitem [{Note1()}]{Note1}%
  \BibitemOpen
  \bibinfo {note} {The authors of Ref.~\cite {kurtosisBalanda1988} refine this
  simplified description to ``kurtosis vaguely [is the] location- and
  scale-free movement of probability mass from the shoulders of a distribution
  into its center and tail.''}\BibitemShut {Stop}%
\bibitem [{Note2()}]{Note2}%
  \BibitemOpen
  \bibinfo {note} {Generating a sample from a Gaussian with two or more clearly
  distinct modes (maxima) is rather improbable, we do not discuss them here. We
  just notice that already bimodal frequency distributions exhibit a rich
  bifurcation map~\cite {PhysRevE.79.026204, PhysRevE.80.046215}.}\BibitemShut
  {Stop}%
\bibitem [{\citenamefont {Petkoski}\ \emph {et~al.}(2013)\citenamefont
  {Petkoski}, \citenamefont {Iatsenko}, \citenamefont {Basnarkov},\ and\
  \citenamefont {Stefanovska}}]{Petkoski2013}%
  \BibitemOpen
  \bibfield  {author} {\bibinfo {author} {\bibfnamefont {S.}~\bibnamefont
  {Petkoski}}, \bibinfo {author} {\bibfnamefont {D.}~\bibnamefont {Iatsenko}},
  \bibinfo {author} {\bibfnamefont {L.}~\bibnamefont {Basnarkov}}, \ and\
  \bibinfo {author} {\bibfnamefont {A.}~\bibnamefont {Stefanovska}},\ }\href
  {\doibase 10.1103/PhysRevE.87.032908} {\bibfield  {journal} {\bibinfo
  {journal} {Phys. Rev. E}\ }\textbf {\bibinfo {volume} {87}},\ \bibinfo
  {pages} {032908} (\bibinfo {year} {2013})}\BibitemShut {NoStop}%
 \bibitem [{\citenamefont {Pearson}(1916)}]{Pearson429}%
  \BibitemOpen
  \bibfield  {author} {\bibinfo {author} {\bibfnamefont {K.}~\bibnamefont
  {Pearson}},\ }\href {\doibase 10.1098/rsta.1916.0009} {\bibfield  {journal}
  {\bibinfo  {journal} {Philosophical Transactions of the Royal Society of
  London A: Mathematical, Physical and Engineering Sciences}\ }\textbf
  {\bibinfo {volume} {216}},\ \bibinfo {pages} {429} (\bibinfo {year}
  {1916})}\BibitemShut {NoStop}%
\bibitem [{\citenamefont {Subbotin}(1923)}]{Sub23}%
  \BibitemOpen
  \bibfield  {author} {\bibinfo {author} {\bibfnamefont {M.}~\bibnamefont
  {Subbotin}},\ }\href {http://mi.mathnet.ru/msb6854} {\bibfield  {journal}
  {\bibinfo  {journal} {Matematicheskii Sbornik}\ }\textbf {\bibinfo {volume}
  {31}},\ \bibinfo {pages} {296} (\bibinfo {year} {1923})}\BibitemShut
  {NoStop}%
\bibitem [{\citenamefont {van Vreeswijk}(1996)}]{vanVreeswijk-96}%
  \BibitemOpen
  \bibfield  {author} {\bibinfo {author} {\bibfnamefont {C.}~\bibnamefont {van
  Vreeswijk}},\ }\href
  {https://journals.aps.org/pre/abstract/10.1103/PhysRevE.54.5522} {\bibfield
  {journal} {\bibinfo  {journal} {Phys. Rev. E}\ }\textbf {\bibinfo {volume}
  {54}},\ \bibinfo {pages} {5522} (\bibinfo {year} {1996})}\BibitemShut
  {NoStop}%
\bibitem [{\citenamefont {Clusella}\ \emph {et~al.}(2016)\citenamefont
  {Clusella}, \citenamefont {Politi},\ and\ \citenamefont
  {Rosenblum}}]{Clusella_etal-16}%
  \BibitemOpen
  \bibfield  {author} {\bibinfo {author} {\bibfnamefont {P.}~\bibnamefont
  {Clusella}}, \bibinfo {author} {\bibfnamefont {A.}~\bibnamefont {Politi}}, \
  and\ \bibinfo {author} {\bibfnamefont {M.}~\bibnamefont {Rosenblum}},\ }\href
  {http://iopscience.iop.org/article/10.1088/1367-2630/18/9/093037/meta}
  {\bibfield  {journal} {\bibinfo  {journal} {New Journal of Physics}\ }\textbf
  {\bibinfo {volume} {18}},\ \bibinfo {pages} {093037} (\bibinfo {year}
  {2016})}\BibitemShut {NoStop}%
\bibitem [{\citenamefont {Balanda}\ and\ \citenamefont
  {MacGillivray}(1988)}]{kurtosisBalanda1988}%
  \BibitemOpen
  \bibfield  {author} {\bibinfo {author} {\bibfnamefont {K.~P.}\ \bibnamefont
  {Balanda}}\ and\ \bibinfo {author} {\bibfnamefont {H.~L.}\ \bibnamefont
  {MacGillivray}},\ }\href {http://www.jstor.org/stable/2684482} {\bibfield
  {journal} {\bibinfo  {journal} {Am. Stat.}\ }\textbf
  {\bibinfo {volume} {42}},\ \bibinfo {pages} {111} (\bibinfo {year}
  {1988})}\BibitemShut {NoStop}%
\bibitem [{\citenamefont {Martens}\ \emph {et~al.}(2009)\citenamefont
  {Martens}, \citenamefont {Barreto}, \citenamefont {Strogatz}, \citenamefont
  {Ott}, \citenamefont {So},\ and\ \citenamefont
  {Antonsen}}]{PhysRevE.79.026204}%
  \BibitemOpen
  \bibfield  {author} {\bibinfo {author} {\bibfnamefont {E.~A.}\ \bibnamefont
  {Martens}}, \bibinfo {author} {\bibfnamefont {E.}~\bibnamefont {Barreto}},
  \bibinfo {author} {\bibfnamefont {S.~H.}\ \bibnamefont {Strogatz}}, \bibinfo
  {author} {\bibfnamefont {E.}~\bibnamefont {Ott}}, \bibinfo {author}
  {\bibfnamefont {P.}~\bibnamefont {So}}, \ and\ \bibinfo {author}
  {\bibfnamefont {T.~M.}\ \bibnamefont {Antonsen}},\ }\href {\doibase
  10.1103/PhysRevE.79.026204} {\bibfield  {journal} {\bibinfo  {journal} {Phys.
  Rev. E}\ }\textbf {\bibinfo {volume} {79}},\ \bibinfo {pages} {026204}
  (\bibinfo {year} {2009})}\BibitemShut {NoStop}%
\end{thebibliography}


\end{document}